\documentclass[prr,aps,twocolumn,10pt,superscriptaddress,notitlepage,nofootinbib,longbibliography]{revtex4-2}
\usepackage[margin=0.75in]{geometry}
\usepackage[caption=false]{subfig}
\usepackage{graphicx}
\usepackage{amsmath,mathtools,bm}
\usepackage{amssymb}
\usepackage{epstopdf}
\usepackage{siunitx}
\usepackage[version=4]{mhchem}
\usepackage{color}
\usepackage[ngerman,english]{babel}
\usepackage{lipsum, babel}
\usepackage{tabularx}
\usepackage{braket}
\usepackage{booktabs} 
\usepackage{appendix}

\definecolor{winered}{rgb}{0.5,0,0}

\DeclareSymbolFont{matha}{OML}{txmi}{m}{it}
\DeclareMathSymbol{\varv}{\mathord}{matha}{118}

\usepackage{amsmath}
\usepackage{graphicx,wrapfig}

\usepackage{amsmath}
\usepackage{float}
\makeatother

\usepackage[utf8]{inputenc}

\usepackage{bibunits}
\defaultbibliographystyle{apsrev4-2}
\defaultbibliography{DiamondOMCs}

\begin{document}

\raggedbottom
\begin{bibunit}

\title{Diamond optomechanical crystals for high-frequency strain and comb generation}

\author{Elham Zohari}
\affiliation{Department of Physics, University of Alberta, Edmonton, AB, T6G 2E1, Canada}
\affiliation{National Research Council of Canada, Quantum and Nanotechnology Research Centre, Edmonton, Alberta, T6G 2M9, Canada}
\affiliation{Institute for Quantum Science and Technology, University of Calgary, Calgary, AB, T2N 1N4, Canada}
\author{Waleed El-Sayed}
\affiliation{Institute for Quantum Science and Technology, University of Calgary, Calgary, AB, T2N 1N4, Canada}
\affiliation{Department of Physics and Astronomy, University of Calgary, Calgary, AB, T2N 1N4, Canada}
\author{Aria Jafari}
\affiliation{Institute for Quantum Science and Technology, University of Calgary, Calgary, AB, T2N 1N4, Canada}
\affiliation{Department of Physics and Astronomy, University of Calgary, Calgary, AB, T2N 1N4, Canada}
\author{Ahmas El-hamamsy}
\affiliation{Institute for Quantum Science and Technology, University of Calgary, Calgary, AB, T2N 1N4, Canada}
\affiliation{Department of Physics and Astronomy, University of Calgary, Calgary, AB, T2N 1N4, Canada}
\author{Peyman Parsa}
\affiliation{Institute for Quantum Science and Technology, University of Calgary, Calgary, AB, T2N 1N4, Canada}
\affiliation{Department of Physics and Astronomy, University of Calgary, Calgary, AB, T2N 1N4, Canada}
\author{Joseph E. Losby}
\affiliation{Institute for Quantum Science and Technology, University of Calgary, Calgary, AB, T2N 1N4, Canada}
\affiliation{Department of Physics and Astronomy, University of Calgary, Calgary, AB, T2N 1N4, Canada}
\author{Natália C. Carvalho}
\affiliation{Institute for Quantum Science and Technology, University of Calgary, Calgary, AB, T2N 1N4, Canada}
\affiliation{Department of Physics and Astronomy, University of Calgary, Calgary, AB, T2N 1N4, Canada}
\author{Paul E. Barclay}
\email[Paul~E.\ Barclay: ]{Corresponding author pbarclay@ucalgary.ca}
\affiliation{Institute for Quantum Science and Technology, University of Calgary, Calgary, AB, T2N 1N4, Canada}
\affiliation{Department of Physics and Astronomy, University of Calgary, Calgary, AB, T2N 1N4, Canada}

\date{\today}

\begin{abstract}
Quantum optomechanical technologies benefit from mechanical oscillators that are high-frequency, can be coherently driven, and are capable of coupling to other quantum systems. Diamond supports all of these criteria: its large elastic modulus increases mechanical resonance frequency, its low nonlinear optical absorption increases the allowed intensity of fields used for coherent optomechanics, and it hosts spin qubits that interact with mechanical modes. Here we demonstrate a diamond optomechanical crystal cavity that supports multiple mechanical resonances with $\sim$\SI{12}{\giga\hertz} frequency and high $Q_\text{m}\times f_\text{m}$ product that can be coherently coupled to multiple optical modes. By exciting this sideband resolved system into mechanical self-sustained oscillations, we generate a frequency comb spanning \SI{143}{\giga\hertz}. Analysis of the comb spectrum, combined with systematic characterization of the system's optomechanical coupling, allows us to quantitatively show that its mechanical oscillation amplitude reaches \SI{130}{\pico\meter}. This corresponds to a maximum total dynamic strain of $1.1\times10^{-3}$ that is sufficiently high for future demonstrations of optomechanical control of diamond spin qubits.
\end{abstract}

\maketitle

\section*{Introduction}

The interaction between light and mechanical motion is central to a growing number of classical and quantum technologies~\cite{Aspelmeyer2014, barzanjeh2022}. In cavity-optomechanical systems, where optical and mechanical resonators interact, light can cool mechanical motion toward the quantum ground state~\cite{Chan2011, riviere2011optomechanical}, enable ultraprecise sensing~\cite{li2021sensing, Krause2012}, and drive mechanical self-oscillations that generates frequency combs~\cite{Grudinin2010, mercade2020microwave, del2007optical}. Optomechanical systems also provide an interface between microwave and optical domains~\cite{andrews2014bidirectional, jiang2020efficient, forsch2020microwave, han2021microwave}, and can be used to process information~\cite{stannigel2012optomechanical}, for example by serving as memories~\cite{lake2021processing} and wavelength converters~\cite{Hill2012}. Many of these capabilities require co-localization of high-quality factor optical and mechanical resonances to sub-wavelength scales to enhance optomechanical coupling. This can be achieved in optomechanical crystal (OMC) cavities~\cite{Eichenfield2009}, which confine GHz-frequency phonons and optical photons to nanoscale volumes.

Optomechanical crystal cavities fabricated from single crystal diamond are an attractive platform for a wide range of applications thanks to the material's exceptional physical properties. Its \SI{5.47}{\eV} bandgap eliminates multiphoton absorption at telecommunication wavelengths, enabling it to support intense optical fields. Its outstanding thermal conductivity (\SI{2200}{\watt\per\meter\per\kelvin}) reduces optical power heating mitigating thermoelastic and Akhiezer damping~\cite{Lifshitz2000TED, ghaffari2013quantum} and thus allowing diamond nanostructures to support ultrahigh quality factor mechanical resonances~\cite{oh2026spin}. Finally, its extreme Young's modulus (\SI{1050}{\giga\pascal})~\cite{diamond_handbook} enhances mechanical resonance frequencies, important to overcome thermal decoherence at higher temperatures, and to achieve sideband-resolved operation, critical for optomechanical backaction~\cite{kim2023diamond, elsayed2026exceptional}. Diamond OMCs can also function as interfaces to solid-state spin qubits formed by defects in their host material~\cite{awschalom2018quantum, doherty2013nitrogen}. Their coupling to mechanical motion via strain provides a phonon-mediated route for spin control~\cite{macquarrie2013, shandilya2021optomechanical, ovartchaiyapong2014dynamic, udvarhelyi2017spin, teissier2014strain}. Realizing strong strain-mediated coupling requires mechanical resonators that concentrate dynamic strain at spin qubit sites~\cite{shandilya2021optomechanical}, and diamond OMC devices~\cite{oh2026spin, raniwala2025spin} are a leading platform for achieving this.

Diamond OMCs have recently been used to demonstrate coherent spin-phonon coupling~\cite{joe2026purcell}, and have been shown to support mechanical quality factors $Q_\text{m}>10^6$ at cryogenic temperatures~\cite{oh2026spin}. Here we demonstrate a diamond OMC that couples multiple telecommunication optical and high-frequency mechanical modes, and characterize its potential for optomechanically driving spin qubits. This device has a desirable combination of high optomechanical coupling, high mechanical frequency $f_\text{m} > 12$ GHz, and room temperature $Q_\text{m}\times f_\text{m}$ product $> 10^{14}$. Operating in the resolved-sideband regime, it exhibits self-sustained nonlinear oscillations in ambient conditions for mW optical input power. By analyzing the resulting optical frequency comb, which consists of optical sidebands spanning \SI{143}{\giga\hertz}, and performing optomechanical spectroscopy to characterize the optomechanical coupling strength of the device's mode spectrum, we quantitatively estimate the maximum self-oscillation amplitude of the device and assess its potential for strain-mediated coupling to diamond spin qubits. We find that a peak mechanical oscillation amplitude of \SI{130}{\pico\meter}, corresponding to the total dynamic strain of $1.1\times10^{-3}$ can be achieved, more than 50 times larger than achieved in previously demonstrated diamond cavity optomechanical systems based on microdisk resonators~\cite{shandilya2021optomechanical}.

\section*{Device Design and Fabrication}

\begin{figure}[h]
    \centering
    \includegraphics[width=\columnwidth]{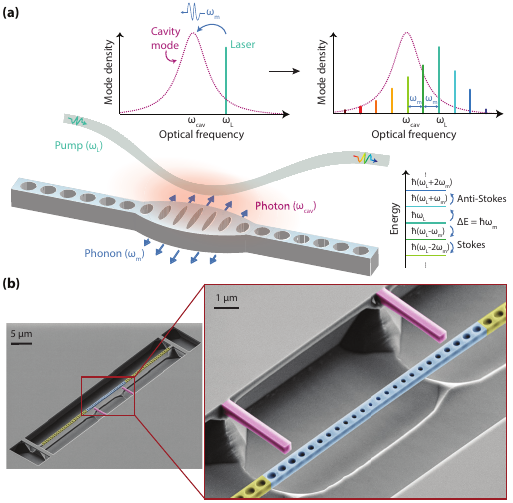}
    \caption{Diamond OMC cavity and cascaded photon–phonon interactions.
    \textbf{(a)} Schematic illustration: A continuous-wave pump laser drives an optomechanical cavity, generating Stokes and anti-Stokes sidebands through cascaded scattering with a mechanical mode at frequency $\omega_\text{m}$. The energy diagram shows the underlying phonon emission and absorption processes, producing equally spaced comb lines at $\omega_\text{L}\pm \omega_\text{m}$.
    \textbf{(b)} False-coloured SEM image of a fabricated diamond OMC cavity, showing the central cavity region (blue), Bragg mirror sections (yellow), and integrated fibre-coupling support structures (pink).}
    \label{fig:illustration}
\end{figure}

The diamond OMC demonstrated here is illustrated in Fig.\ \ref{fig:illustration}(a), and consists of a suspended diamond nanobeam patterned with an array of elliptical air holes. Scanning electron microscope (SEM) images of the fabricated device are shown in Fig.\ \ref{fig:illustration}(b). It was created from a single-crystal diamond chip (Element Six, optical grade CVD grown) using the SCREAM~\cite{SHAW1994} inspired quasi-isotropic diamond undercut etching process demonstrated in Refs.\ \cite{Khanaliloo2015MD, Khanaliloo2015NB}. The process was modified, following Refs.\ \cite{Mouradian2017, wan2018two}, to incorporate \ce{Al2O3} in place of \ce{Si3N4} for the sidewall protection step.

The device was designed using photonic-phononic co-design principles~\cite{Safavi2010,Chan2012}. A defect within the periodic array of holes was created by gradually varying the hole ellipticity and lattice constant. The resulting structure localizes both photons and phonons within the defect region at frequencies within the respective optical and mechanical bandgaps of the mirror region unit cells. The taper profile and nanobeam dimensions (width \SI{577.5}{\nano\meter} and thickness \SI{305}{\nano\meter}) were optimized using finite element method simulations (COMSOL Multiphysics)~\cite{moraes2022optimization} to maximize the photon-phonon vacuum optomechanical coupling rate, $g_{0}$, while maintaining high optical ($Q_\text{opt}$) and mechanical ($Q_\text{m}$) quality factors (see Supplemental Material for the full design details). 

\begin{figure*}
    \centering
    \includegraphics[width=\textwidth]{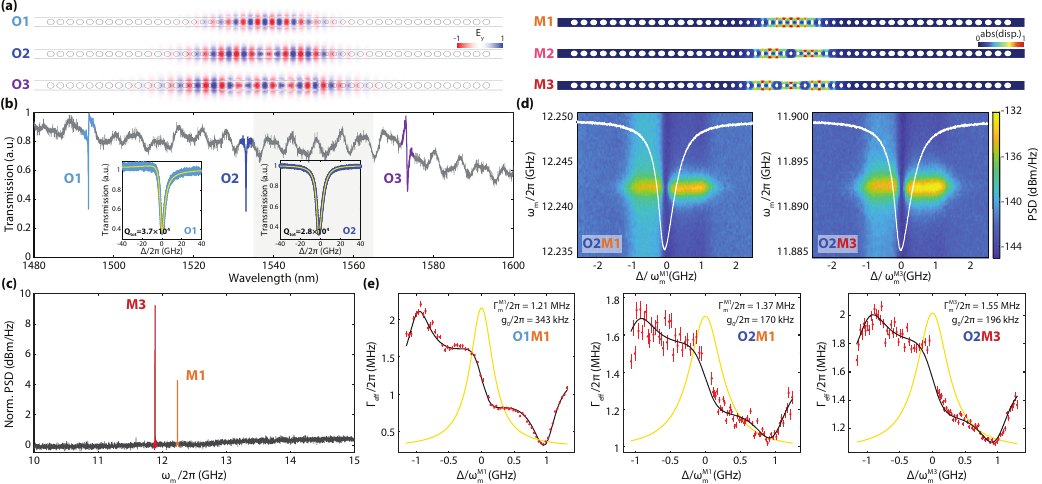}
    \caption{Diamond OMC cavity mode characterization.
    \textbf{(a)} Simulated spatial profiles of the optical and mechanical cavity modes. Left: Electric field intensity distributions of the first- (O1, \SI{1497}{\nano\meter}), second- (O2, \SI{1530}{\nano\meter}), and third-order (O3, \SI{1566}{\nano\meter}) optical modes. Right: Mechanical displacement profiles of the first- (M1, \SI{12.43}{\giga\hertz}), second- (M2, \SI{12.27}{\giga\hertz}), and third-order (M3, \SI{12.08}{\giga\hertz}) acoustic breathing modes. The colour scheme identifying each mode
    is used consistently throughout.
    \textbf{(b)} Normalized broadband transmission spectrum of the diamond OMC showing three optical resonances at 1494, 1530, and \SI{1571}{\nano\meter}. The shaded region shows the operational range of the EDFA used. Insets show high-resolution scans of O1 and O2 fit (yellow curve). 
    \textbf{(c)} Broadband normalized RF spectra revealing acoustic modes M1 at \SI{12.24}{\giga\hertz} and M3 at \SI{11.89}{\giga\hertz}.
    \textbf{(d)} Power spectral density (PSD) of the M1 and M3 mechanical modes as a function of normalized laser detuning ($\Delta/\omega_\text{m}^\text{M1}$ and $\Delta/\omega_\text{m}^\text{M3}$), overlaid with the optical transmission (white curve), demonstrating optomechanical transduction across the cavity linewidth.
    \textbf{(e)} Optomechanical backaction for three mode pairs with the highest $g_0$, O1--M1, O2--M1, and O2--M3 with fits to the data (black curves) yielding 
    coupling rates $g_0/2\pi =$343, 170, and \SI{196}{\kilo\hertz}, respectively. The yellow curve shows the intracavity photon density in arbitrary units.}
    \label{fig:OptomechMeasure}
\end{figure*}

The resulting transverse-electric (TE-like) optical modes (O1, O2, O3) have simulated wavelengths between 1497 -- 1566~nm. Due to the narrow width of the optimized device, vertical radiation loss limits the theoretically predicted optical quality factor $Q_\text{opt} < 1.5 \times 10^5$~\cite{quan2011}. This small width serves to enhance the frequencies $f_\text{m} = [12.43, 12.27, 12.08]$~GHz of three localized mechanical breathing modes (M1, M2, M3), whose simulated spatial profiles are shown in Fig.\ \ref{fig:OptomechMeasure}(a), and whose effective mass are $m_\text{eff} = [0.55, 0.63, 0.69]$~fg. All of these modes have simulated mechanical quality factor $Q_\text{m} > 10^6$. The optomechanical coupling rate between these modes and the optical modes, considering both moving boundaries and the photoelastic effect, was calculated following Ref.~\cite{moraes2022optimization}. The mode combination with the highest single-photon optomechanical coupling rate, $g_0/2\pi = \SI{343}{\kilo\hertz}$, is between the fundamental mechanical breathing mode (M1) and the fundamental optical mode (O1). This $g_0$ exceeds previously demonstrated optomechanical coupling rates in diamond OMCs operating at telecommunication wavelengths \cite{Burek2016, elsayed2026exceptional}. The second-order mechanical breathing mode (M2) exhibits vanishing coupling to all optical modes due to its odd spatial symmetry. The third-order breathing mode (M3), couples most strongly to the second-order optical mode (O2), with simulated $g_0/2\pi=\SI{198}{\kilo\hertz}$. The O2 mode, which also couples well to the M1 breathing mode, lies within the C-band optical amplifier range used in our experimental setup, making the O2--M3 pair the natural operating point for this work. All simulated device parameters are summarized in Table~\ref{tab:params}.

\begin{table*}
\centering
\caption{Diamond OMC performance parameters: simulated and measured (sim., meas.) optomechanical coupling rate $g_0/2\pi$ with its moving-boundary (MB) and photoelastic (PE) contributions; simulated and measured optical quality factor $Q_\text{opt}$; measured intrinsic and total cavity intensity decay rates $\kappa_\text{int}$, $\kappa_\text{tot}$; mechanical quality factor $Q_\text{m}$; and single-photon cooperativity $C_0=4g_0^2/\kappa_\text{tot}\Gamma_\text{m}$, calculated from measured cavity parameters. M2 mode is omitted due to its vanishing coupling to all optical modes.}
\label{tab:params}
\resizebox{\textwidth}{!}{
\begin{tabular}{l cc cc cc cc ccc cc cc cc} 
\toprule
& \multicolumn{2}{c}{$g_\text{MB}/2\pi$ (kHz)} 
& \multicolumn{2}{c}{$g_\text{PE}/2\pi$ (kHz)} 
& \multicolumn{2}{c}{$g_0/2\pi$ (kHz)} 
& \multicolumn{2}{c}{Meas. $g_0/2\pi$ (kHz)} 
& \multicolumn{3}{c}{$Q_\text{opt}\ (\times 10^{4})$} 
& \multicolumn{2}{c}{$\kappa/2\pi$ (GHZ)}
& \multicolumn{2}{c}{$Q_\text{m}\ (\times 10^{3})$}
& \multicolumn{2}{c}{$C_0\ (\times 10^{-7})$} \\
\cmidrule(lr){2-3}\cmidrule(lr){4-5}\cmidrule(lr){6-7}\cmidrule(lr){8-9}\cmidrule(lr){10-12}\cmidrule(lr){13-14}\cmidrule(lr){15-16}\cmidrule(lr){17-18}
& {M1} & {M3} & {M1} & {M3} & {M1} & {M3} & {M1} & {M3} & {Sim.} & {Meas. int.} & {Meas. tot.} & {$\kappa_\text{int}/2\pi$} & {$\kappa_\text{tot}/2\pi$} & {M1} & {M3} & {M1} & {M3} \\
\midrule
\textbf{O1} & 127 & 19 & 216 & 32 & 343 & 51 & 343(2) & \text{--} & 15 & 6.3(1) & 3.74(5) & 3.18(7) & 5.39(7) & 10.12(6) & \text{--} & 720(10) & \text{--} \\
\textbf{O2} & 63 & 74 & 100 & 124 & 163 & 198 & 170(2) & 196(1) & 7.6 & 4.59(1) & 2.755(6) & 4.28(1) & 7.11(2) & 8.94(4) & 7.67(2) & 119(3) & 139(2) \\
\textbf{O3} & 51 & 19 & 76 & 29 & 127 & 48 & \text{--} & \text{--} & 3.9 & 2.69(1) & 2.58(1) & 7.09(3) & 7.40(2) & \text{--} & \text{--} & \text{--} & \text{--} \\
\bottomrule
\end{tabular}
}
\end{table*}

\section*{Optomechanical Coupling Characterization}

The OMCs were characterized using a dimpled fibre taper probe~\cite{michael2007} in ambient conditions. Light from a tunable laser was amplified with an erbium-doped fibre amplifier (EDFA) and coupled into and out of the devices positioned in the fibre taper near field. The transmitted field was monitored with a photodetector whose DC output monitors the steady-state optical cavity response, and whose RF output was connected to a real-time spectrum analyzer (RSA), allowing fluctuations in the optical field due to resonator motion to be measured. Figure\ \ref{fig:OptomechMeasure}(b) shows the normalized transmission when the input laser was swept over its full tuning range, revealing three resonances corresponding to the cavity modes O1, O2 and O3 near [1491, 1530, 1571]~nm with loaded quality factors $Q_\text{opt} = [3.7, 2.8, 2.6]\times10^{4}$. These values, as well as the corresponding loaded ($\kappa_\text{tot}$) and intrinsic ($\kappa_\text{int}$) cavity mode energy decay rates extracted from the fits to the resonances, are summarized in Table~\ref{tab:params}.

The OMC's optical modes allow motion of its mechanical resonances to be transduced by the transmitted optical field, as shown in Fig.\ \ref{fig:OptomechMeasure}(c,d), which plot the photodetected RF power spectral density (PSD) when the laser frequency $\omega_\text{L}$ is blue-detuned ($\Delta=\omega_\text{L}-\omega_\text{cav}^{\text{O2}} > 0$) and swept across the O2 resonance. Two prominent peaks are visible at $\omega_\text{m}^\text{M1}/2\pi = 12.24\,\text{GHz}$ and $\omega_\text{m}^\text{M3}/2\pi = 11.89\,\text{GHz}$ generated by thermal motion of the device's mechanical breathing modes M1 and M3, respectively. The measured frequency values are in good agreement with simulated mechanical frequencies given above, and their spacing of 350~MHz is nearly identical to theoretical predictions. The combination of $\kappa_\text{tot}$ and $\omega_\text{m}$ measured here places the device in the sideband-resolved regime ($\omega_\text{m}>\kappa_\text{tot}$), a prerequisite for ground-state cooling and coherent quantum state manipulation~\cite{marquardt2007}.

To characterize the device's optomechanical coupling, we analyze optomechanical backaction-induced modification to the thermomechanical linewidth, $\Gamma_\text{eff}$,
as $\Delta$ is varied. In Fig.\ \ref{fig:OptomechMeasure}(e) we plot the measured linewidth $\Gamma_\text{eff}(\Delta)$ of the M1 mode for varying detuning from the O1 and O2 resonances, and for the M3 mode for varying detuning from the O2 resonance. As expected, the device exhibits optomechanical damping (anti-damping) for $\Delta < 0$ ($\Delta >0$). Also shown are the theoretical fits, obtained from $\Gamma_\text{eff} = \Gamma_\text{m} + \Gamma_\text{om}(\Delta)$, where $\Gamma_\text{om} = n g_0^2 \left(\frac{\kappa_\text{tot}}{(\Delta + \omega_\text{m})^2 + (\kappa_\text{tot}/2)^2} - \frac{\kappa_\text{tot}}{(\Delta - \omega_\text{m})^2 + (\kappa_\text{tot}/2)^2}\right)$ is the optomechanical backaction induced change in mechanical resonance intrinsic linewidth $\Gamma_\text{m}$. Here $n = n(\Delta)$ is the intracavity photon number determined by the cavity-fibre optical coupling and corresponding resonance constrast, and the input power $P_\text{i} = [1.22, 1.75, 1.74]$~mW for mode pairs [O1--M1, O2--M1, O2--M3], respectively. From these fits we extract $g_0/2\pi=[343, 170, 196]$~kHz. The excellent agreement for different mode pairs with the simulated values listed in Table\ \ref{tab:params}, validates the FEM simulation methodology and, importantly, underpins the quantitative estimation of resonator displacement and strain later in this work. We note that the peak intracavity photon number $n_\text{max} \sim 2.5\times10^5$ in these measurements is achieved without significant thermo-optic heating in part owing to diamond's aforementioned low nonlinear absorption and high thermal conductivity.

The measured intrinsic mechanical linewidths of $\Gamma_\text{m}/2\pi= [1.2, 1.4, 1.6]$~MHz, correspond to mechanical quality factors of $Q_\text{m}= [10, 8.9, 7.7]\times10^3$, for mode pairs [O1--M1, O2--M1, O2--M3], respectively. These mechanical quality factors are consistent with values reported for diamond OMC devices measured at room temperature~\cite{Burek2016, Joe2024, cady2019diamond}, suggesting that surface and phonon-phonon scattering losses are dominant room-temperature dissipation channels for this material and geometry. Note that the M1 mode possesses $Q_\text{m} \times f_\text{m} = 1.2\times10^{14}$~Hz, exceeding the room temperature threshold of coherent operation by a factor of 20~\cite{norte2016mechanical, khosla2017quantum, tsaturyan2017ultracoherent} and that of previous room-temperature nanophotonic cavity-optomechanical systems by a factor of 3.4~\cite{Ren2020}.

\section*{Frequency Comb Generation}

The high mechanical frequencies of diamond OMCs make them an attractive platform for generating microwave tones. This can be achieved by increasing $P_\text{i}$ of an input laser blue-detuned to the Stokes frequency ($\Delta = \omega_\text{m}$), where it can optomechanically amplify the resonator motion. If optomechanical backaction overcomes intrinsic mechanical damping ($\Gamma_\text{eff} \le 0$)~\cite{Marquardt2006, Kippenberg2008}, the mechanical resonance enters self-sustained oscillations~\cite{Carmon2005, Rokhsari2002}. Here we demonstrate self-oscillation of the M3 mode and characterize the resulting optical frequency comb created by nonlinear optical transduction of the resonator motion, which extends over a \SI{143}{\giga\hertz} range.

\begin{figure}[h]
    \centering
    \includegraphics[width=\columnwidth]{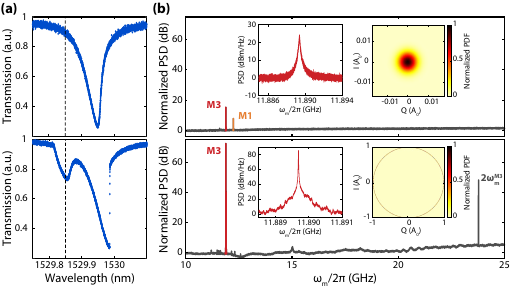}
    \caption{Thermal-to-self-oscillation transition of the M3 mechanical mode.
    \textbf{(a)} Normalized optical transmission at input powers $P_\text{i} = 7.1$~mW (top) and $19.1$~mW (bottom). The dashed black line indicates the laser detuning ($\Delta/\omega_\text{m}^\text{M3}=1.0$) used for all measurements.
    \textbf{(b)} Normalized PSD of the optically transduced RF spectrum, showing thermal Brownian motion (top) and self-sustained oscillation (bottom). Insets: high-resolution PSD of the M3 mode showing linewidth narrowing at self-oscillation, and phase-space distributions from IQ quadrature measurements showing the PDF for thermal noise and limit-cycle oscillation.}
    \label{fig:mechanics}
\end{figure}

Figure\ \ref{fig:mechanics}(a) shows the fibre taper transmission near the O2 mode at input powers below and above the self-oscillation threshold, $P_\text{i} = [7.1, 19.1]$~mW, respectively, corresponding to $n = [1.0, 9.4]\times10^5$ intracavity photons when $\Delta = \omega_\text{m}^\text{M3}$. The mechanical mode's transition from incoherent thermal motion to large amplitude self-oscillation can be observed in both the optical transmission spectrum and the mechanical PSD. For low $P_\text{i}$, the optical transmission exhibits a Lorentzian lineshape (Fig.\ \ref{fig:mechanics}(a) top). When the optical power is increased above the self-oscillation threshold, large amplitude oscillations modulate the instantaneous optical transmission. This creates shoulders on the blue-detuned region of the time averaged transmission spectrum where self-oscillations are generated (Fig.\ \ref{fig:mechanics}(a) bottom)~\cite{Carmon2005, Schliesser2008}. Simultaneously, as shown in Fig.\ \ref{fig:mechanics}(b), the PSD of the M3 mode narrows and increases in amplitude by 50~dB. The phase space representation of the photodetected signal, also shown in Fig.\ \ref{fig:mechanics}(b), transforms from a broad Gaussian characteristic of thermomechanical motion to a sharp ring characteristic of coherent oscillations at the mechanical frequency. These phase-space plots show the probability density distributions of the in-phase (I) and quadrature (Q) components of the mechanical displacement transduced by the photoreceiver~\cite{Clarke2023}. In the self-oscillation regime, the sharp ring has a radius proportional to the mechanical oscillation amplitude $A$, while the phase undergoes slow diffusion~\cite{weiss2016noise, catalini2021modeling}.

The coherent modulation of the cavity field by the self-oscillating resonator also generates harmonic sidebands. The second order ($2\omega_\text{m}^\text{M3}$) sideband is visible in the PSD measurement in Fig.\ \ref{fig:mechanics}(b), however higher order sidebands fall outside the bandwidth of the detection electronics. We instead measure them with an optical spectrum analyzer (OSA - APEX) that uses heterodyne detection to achieve \SI{5}{\mega\hertz} resolution over the full optical bandwidth of the resonance.

\begin{figure*}
    \centering
    \includegraphics[width=\textwidth]{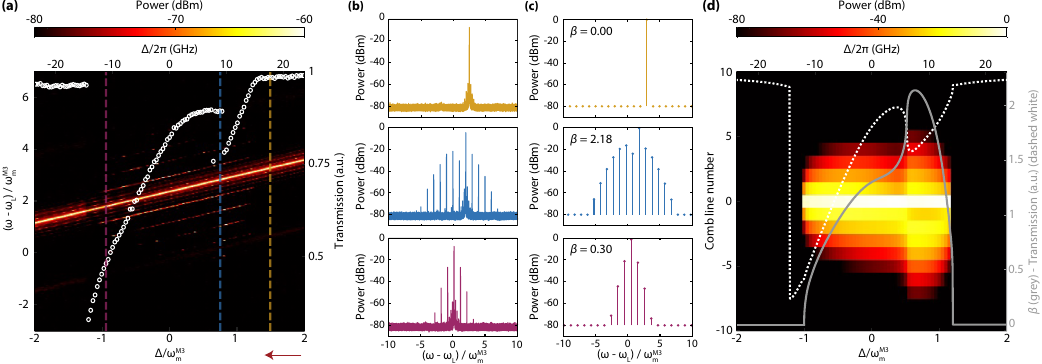}
    \caption{Cascaded photon-phonon scattering and optical frequency comb generation. (a) Two-dimensional map of the transmitted optical spectrum as a function of normalized laser-cavity detuning $\Delta/\omega_\text{m}^\text{M3}$, recorded by sweeping the laser across the O2 resonance (red arrow indicates sweep direction). The overlaid white circles indicate the cavity transmission profile. (The colour scale is truncated to enhance visibility of the sideband structure.) (b) Optical spectra extracted at the three detunings marked in (a), showing the evolution of the optomechanical frequency comb. At optimal blue detuning, thirteen equally spaced comb lines appear spanning \SI{143}{\giga\hertz}. The comb spacing matches the M3 mechanical frequency $\omega_\text{m}^\text{M3}/2\pi = $\SI{11.89}{\giga\hertz}. (c) Simulated optical spectra at the same detunings, using measured device parameters, with modulation index varying from $\beta = 0$ to $\beta_\mathrm{max} = 2.18$ across the detuning sweep. (d) Simulated optical comb spectra versus normalized laser detuning. White dashed line: fit to the optical cavity transmission profile shown in panel (a). Grey solid line: modulation index $\beta$ (right axis).}
    \label{fig:OSA_Comb}
\end{figure*}

Figure\ \ref{fig:OSA_Comb}(a) shows the transmission lineshape measured for $P_\text{i} = 22.7$~mW (${n} = 1.3 \times10^6$, dropped power $P_\text{d} = 5.4$~mW when $\Delta \sim \omega_\text{m}^\text{M3}$) when $\Delta$ is scanned from blue- to red-detuned, overlaid upon a spectrograph measured by the OSA. As the input field is tuned across the O2 cavity mode, a peak corresponding to the laser frequency becomes flanked by sidebands spaced by $\omega_\text{m}^\text{M3}$. Figure\ \ref{fig:OSA_Comb}(b) shows slices from the spectrograph at representative detunings indicated by the dashed lines in Fig.\ \ref{fig:OSA_Comb}(a), highlighting the frequency comb structure. The broadest comb spectrum occurs in the shoulder of the transmission lineshape, where as highlighted in Fig.\ \ref{fig:OSA_Comb}(b) thirteen comb lines are observed spanning \SI{143}{\giga\hertz} with spacing of $\omega_\text{m}^\text{M3}/2\pi = \SI{11.89}{\giga\hertz}$. The asymmetry of seven red-shifted and five blue-shifted sidebands relative to the carrier arises due to the red-shifted (Stokes) sidebands falling closer to the cavity resonance, while blue-shifted sidebands (anti-Stokes) move progressively further from cavity resonance. As $\Delta$ is tuned away from this optimal point, the number of sidebands decreases, reflecting the reduction in mechanical oscillation amplitude and modulation depth.

In these measurements the cavity lineshape is distorted into a shark-fin shape by photothermal effects commonly observed at sufficiently high power in nanoscale cavities, such that the cavity resonance frequency $\omega_\text{cav}^0$ shifts as the input laser is scanned across the resonance. We therefore define the effective, instantaneous detuning $\Delta_\text{eff} \equiv \omega_\text{L} - \omega_\text{cav}^\text{0} = \Delta - \delta\omega_\text{PT}$ which is shifted by $\delta\omega_\text{PT}$ due to the $n$ dependent photothermal effect and governs the intracavity dynamics, and account for this distinction in modelling the $\Delta$-dependent frequency comb spectra, as discussed below.

The observed frequency combs spectra can be modelled by solving the nonlinear cavity optomechanical equations of motion~\cite{Marquardt2006, parsa2026large} to calculate the self-oscillation amplitude $A$. Given the resulting time dependent modulation of the cavity resonance frequency, $\omega_\text{cav}(t) = \omega_\text{cav}^0 - GA\cos(\omega_\text{m} t)$, governed by optomechanical coupling coefficient $G = \partial\omega_\text{cav}/\partial x$, the intracavity field amplitude satisfies the well known ansatz $\alpha(t) = e^{i\beta\sin(\omega t)}\sum_{m=-\infty}^{\infty}\alpha_m\,e^{-in\omega t}$~\cite{Marquardt2006}. Here the modulation index, $\beta = GA/\omega_\text{m}$, together with the cavity lineshape, determines the sideband amplitudes,
\begin{equation}
    \alpha_m = \frac{\sqrt{\kappa_\text{ex}}\,\alpha_\text{in}\,J_m(\beta)}{\kappa_\text{tot}/2 - i(\Delta_\text{eff} + m \omega_\text{m})}
    \label{eq:alpha_n}
\end{equation}
where $J_m$ are the Bessel functions of the first kind. Note that the cavity susceptibility $\chi(m) \equiv \left[\kappa_\text{tot}/2 - i(\Delta_\text{eff} + n\omega_\text{m})\right]^{-1}$ filters sidebands of order $m$ according to its spectral proximity to the optical cavity resonance. Here $|\alpha_\text{in}|^2 = P_\text{in}$ and $\kappa_\text{ex}$ is the loss rate of the cavity into the transmitted fibre taper mode.
The power output in the $m$-th comb line, $S_m$, is given by
\begin{equation}
    S_m = \left| \sum_{l=-\infty}^{\infty} J_{m-l}(\beta)\,J_l (\beta)\,t(l)\right|^2
    \label{eq:comb}
\end{equation}
where $t(l) = 1 - \kappa_\mathrm{ex}\,\chi(l)$. For large amplitude oscillations, $\beta \gtrsim 1$, the observed frequency comb emerges. Note that direct coupling between non-adjacent sideband orders $l$ and $m$ is suppressed by $(g_0/\omega_\mathrm{m})^{|l-m|} \ll 1$ for $|l-m|>1$~\cite{Parsa2023}. Since $g_0/\omega_\text{m}^\text{M3} = 1.6\times10^{-5}$ for the O2--M3 mode pair, even two-phonon direct transitions between next-nearest sidebands are suppressed by a factor of $\sim10^{-10}$, indicating that the comb builds up through sequential scattering processes.

Figure\ \ref{fig:OSA_Comb}(c) shows combs spectra predicted from this model, calculated at the same representative detunings as the measurements in Fig.\ \ref{fig:OSA_Comb}(b). The predictions are in excellent agreement with the experimental data, including the number of sidebands, the sideband asymmetry, and the suppression of $S_{-1}$ when $\Delta_\text{eff} = \omega_\text{m}^\text{M3}$. Here we followed Ref.\ \cite{parsa2026large} to first solve for $\beta(\Delta)$ from the nonlinear equations of motion, taking into account the thermo-optic effect using a fitting parameter $\zeta$ governing the resulting cavity shift $\delta\omega_\text{PT} = \zeta {n}(\Delta)$ and chosen so that the predicted cavity transmission lineshape matches the measured average transmission. We assumed that the thermal timescale is slow compared to the mechanical frequency, as previously shown in silicon OMC devices~\cite{meenehan2015pulsed, krause2015nonlinear}, so that thermal dynamics can be neglected. The good agreement between theory and experiment is further illustrated in Fig.\ \ref{fig:OSA_Comb}(d), which shows the predicted optical spectrograph across the full $\Delta$ sweep. Also shown is $\beta(\Delta)$, which peaks at $\beta_\text{max} = 2.18$, and the predicted cavity transmission profile, which agrees well with the measured lineshape. From this fitting, we extract a maximum cooperativity $C_\text{om} = 4 n g_0^2/\kappa_\text{tot}\Gamma_\text{m} > 18.1$, showing that thanks to possessing excellent optomechanical and thermal properties, the diamond OMCs allow highly coherent optomechanical coupling. Operating with the O1-M1 mode combination would further enhance these interaction owing to their higher $g_0$.

\section*{Discussion}

The strong agreement between our model and both the transmission lineshape and the frequency comb spectra allows us to combine $\beta(\Delta)$ with the extracted $g_0$ to determine the oscillation amplitude. From $G = g_0/x_\text{zpf}$, where $x_\text{zpf}^\text{M3} =\sqrt{\hbar/2m_\text{eff}\omega_\text{m}} = 1.0$~fm is the zero point fluctuation amplitude for the M3 mode, the maximum self-oscillation amplitude is $A_\text{max} = {\beta_\text{max}\,\omega_\text{m}^\text{M3}}/{G} = 130$~pm. 

Using $A_{\text{max}}$ we can assess the potential for coupling the corresponding internal dynamic strain field to spin and orbital states of colour centres hosted in diamond. From finite element simulations, we find the maximum magnitude of total strain per phonon of M3 to be $\epsilon_\text{zpf} = 9.3 \times 10^{-9}$, located at the edge of the central cavity hole. The observed self-oscillations generate dynamic strain with amplitude $\epsilon = \epsilon_\text{zpf} \times A_\text{max} / x_\text{zpf}= 1.1 \times 10^{-3}$. This is approximately an order of magnitude larger than that used in Ref.\ \cite{macquarrie2015coherent} for coherently manipulating NV centre spins.

The predicted strain coupling in our device could also be used at the single phonon level to couple to SiV colour centres. Following the methodology of Raniwala et al.\ \cite{raniwala2025spin}, we estimate the spin-mechanical coupling rate by projecting the simulated strain tensor onto the SiV's transverse strain component for a representative [111]-oriented SiV defect, with the OMC nanobeam aligned along the diamond [100] axis~\cite{elsayed2026exceptional}. This calculation represents an upper bound, since the realized coupling also depends on the local static strain environment and any applied magnetic field. At the strain maximum near the central cavity hole, we obtain $g_\mathrm{sm}/2\pi = \SI{5.61}{\mega\hertz}$, exceeding demonstrated thresholds for coherent manipulation of diamond colour-centre spins~\cite{udvarhelyi2017spin, doherty2013nitrogen} and establishing this mechanical mode as strain-relevant for spin-based quantum control.

The optical frequency comb demonstrated here has, to our knowledge, both larger comb tooth spacing (\SI{12}{\giga\hertz}) and total bandwidth (\SI{143}{\giga\hertz}) than previously reported optomechanical systems~\cite{Hu2021,gou2025chip,Wang2024,Ng2023, mercade2020microwave, wang2024optomechanical, wan2025optomechanical}. Optomechanical frequency combs offer advantages over conventional nonlinear optical frequency comb platforms, namely that the comb spacing is set by the mechanical resonance frequency vis-à-vis the device geometry and material parameters. Furthermore, the \SI{12}{\giga\hertz} spacing demonstrated here would typically require a much larger mm-scale device if implemented using a purely nonlinear optical process reliant upon optical modes spaced by the target comb spacing. However, we emphasize that nonlinear optical platforms can achieve frequency comb bandwidths far exceeding what is achieved here. 
 
\section*{Conclusion}

We have demonstrated a single-crystal diamond optomechanical crystal cavity exhibiting state-of-the-art vacuum optomechanical coupling rates at telecommunication wavelengths of $g_0/2\pi = \SI{343}{\kilo\hertz}$. Combined with optical quality factors exceeding $Q_{\text{int}} = 6.3 \times 10^4$, and high frequency 12~GHz mechanical modes that are sideband resolved, the device supports self-sustained mechanical oscillations at mW continuous-wave optical power, generating an optical frequency comb whose bandwidth of 143 GHz advances limits for these compact source. The $Q_\text{m}\times f_\text{m}=1.2\times10^{14}$~Hz value achieved here is, to our knowledge, the largest reported for a GHz-frequency optomechanical cavity fabricated from diamond~\cite{Joe2024,Burek2016,cady2019diamond} or other materials~\cite{Ren2020} when operating in ambient condition, and exceeds that of some devices operating in cryogenic conditions~\cite{Chan2011, krause2015nonlinear}. Future studies at mK temperatures, combined with devices fabricated from electronic-grade single-crystal diamond and incorporating phononic shielding are required to fully assess its mechanical coherence relative to other state-of-the-art OMC demonstrations~\cite{Ren2020, meenehan2015pulsed, maccabe2020nano, Joe2024, oh2026spin, li2024ultracoherent} and implement experiments in quantum optomechanics. Diamond's resilience to optical heating makes the device studied here promising for implementing optomechanical quantum memory~\cite{wallucks2020quantum,kristensen2024long} and the development of quantum transducers~\cite{Stannigel2010transduction, sonar2025high}. Experiments integrating nitrogen-vacancy or silicon-vacancy centres with these high-frequency mechanical modes are a natural next step, building on the strain-mediated coupling rates established here, toward demonstrating coherent spin-phonon coupling and optomechanical control of quantum emitters. The ability of this device to support multiple optical and mechanical modes with comparable optomechanical coupling is also promising for implementing experiments in multimode optomechanics.

\putbib[main]
\end{bibunit}

\clearpage
\begin{bibunit}
\onecolumngrid
\appendix
\begin{center}
{\large\bfseries Supplementary Information}
\end{center}

\subsection*{Design and Simulation of the Diamond optomechanical crystal Cavity}

The device is a suspended, single-crystal diamond nanobeam with a periodic elliptical-hole lattice, creating simultaneous photonic and phononic band gaps~\cite{Safavi2010}. A defect cavity is formed by tapering the lattice constant and hole dimensions from the mirror lattice to the cavity unit cell~\cite{Chan2011}. This design co-localizes telecom-wavelength ($\sim$1550~nm) optical modes with $\sim$13~GHz mechanical breathing modes, enabling large vacuum optomechanical coupling rates $g_0$. Relative to our earlier diamond OMCs~\cite{Zohari2022, elsayed2026exceptional}, the present design was optimized to support multimode operation and higher mechanical frequencies $\omega_\text{m}$. 

\begin{figure}[h]
    \centering
    \includegraphics[width = \textwidth]{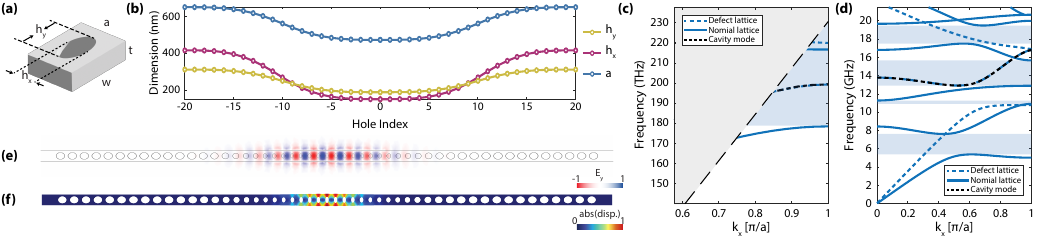}
    \caption{Device design parameters and simulated mode profiles for the 1D OMC. (a) Unit cell schematic showing $a$, $h_\text{x}$, $h_\text{y}$, $w$, and $t$. (b) Evolution of unit cell parameters through the tapered defect region. Simulated (c) optical and (d) mechanical band diagrams. Simulated fundamental (e) optical mode profile (electric field intensity) and (f) mechanical breathing-mode profile (displacement magnitude).}
    \label{fig:design_schematic}
\end{figure}

The narrow beam width required to realize large $\omega_\text{m}$ necessitates a larger lattice constant $a$ to simultaneously support optical modes at telecom wavelengths. This shifts the optical modes closer to the light line, increasing their susceptibility to radiative loss. To mitigate this loss, the unit cell parameters $w$, $t$, $a$, $h_x$, and $h_y$ (Fig.~\ref{fig:design_schematic}) are varied gradually to form an extended cavity region in which the lattice holes remain nearly identical. The resulting long, weakly perturbed cavity minimizes optical radiation loss while simultaneously supporting multiple confined optical and mechanical modes.

The nearly flat parameter profile across the central cavity region gives rise to multiple modes that are closely spaced in frequency and strongly co-localized in space. As a result, several mechanical and optical modes exhibit substantial simultaneous spatial overlap, enabling appreciable optomechanical coupling across multiple optical--mechanical mode pairs.

Mode frequencies, $Q$, and $g_0$ were obtained from FEM simulation (COMSOL Multiphysics), with the taper profile and mirror cell count optimized in MATLAB to maximize $g_0$ at high $Q$. The nominal design couples TE-like optical modes near 1550~nm to a $\sim$ 13~GHz mechanical breathing mode. To explore a broader design space, devices spanning a range of geometries around this nominal design were fabricated in parallel; the device reported in the main text was selected based on its performance and its relevance to the goals of this study. The simulated $Q$ and $g_0$ values reported in the main text correspond to this selected device's specific geometry, rather than to the nominal design.

The device also incorporates fibre-support landing pads, which fix the position of the tapered fibre and reduce drift caused by electrostatic fibre--device attraction at high optical power.

\putbib[main]

\end{bibunit}


\begin{thebibliography}{75}%
\makeatletter
\providecommand \@ifxundefined [1]{%
 \@ifx{#1\undefined}
}%
\providecommand \@ifnum [1]{%
 \ifnum #1\expandafter \@firstoftwo
 \else \expandafter \@secondoftwo
 \fi
}%
\providecommand \@ifx [1]{%
 \ifx #1\expandafter \@firstoftwo
 \else \expandafter \@secondoftwo
 \fi
}%
\providecommand \natexlab [1]{#1}%
\providecommand \enquote  [1]{``#1''}%
\providecommand \bibnamefont  [1]{#1}%
\providecommand \bibfnamefont [1]{#1}%
\providecommand \citenamefont [1]{#1}%
\providecommand \href@noop [0]{\@secondoftwo}%
\providecommand \href [0]{\begingroup \@sanitize@url \@href}%
\providecommand \@href[1]{\@@startlink{#1}\@@href}%
\providecommand \@@href[1]{\endgroup#1\@@endlink}%
\providecommand \@sanitize@url [0]{\catcode `\\12\catcode `\$12\catcode `\&12\catcode `\#12\catcode `\^12\catcode `\_12\catcode `\%12\relax}%
\providecommand \@@startlink[1]{}%
\providecommand \@@endlink[0]{}%
\providecommand \url  [0]{\begingroup\@sanitize@url \@url }%
\providecommand \@url [1]{\endgroup\@href {#1}{\urlprefix }}%
\providecommand \urlprefix  [0]{URL }%
\providecommand \Eprint [0]{\href }%
\providecommand \doibase [0]{https://doi.org/}%
\providecommand \selectlanguage [0]{\@gobble}%
\providecommand \bibinfo  [0]{\@secondoftwo}%
\providecommand \bibfield  [0]{\@secondoftwo}%
\providecommand \translation [1]{[#1]}%
\providecommand \BibitemOpen [0]{}%
\providecommand \bibitemStop [0]{}%
\providecommand \bibitemNoStop [0]{.\EOS\space}%
\providecommand \EOS [0]{\spacefactor3000\relax}%
\providecommand \BibitemShut  [1]{\csname bibitem#1\endcsname}%
\let\auto@bib@innerbib\@empty
\bibitem [{\citenamefont {Aspelmeyer}\ \emph {et~al.}(2014)\citenamefont {Aspelmeyer}, \citenamefont {Kippenberg},\ and\ \citenamefont {Marquardt}}]{Aspelmeyer2014}%
  \BibitemOpen
  \bibfield  {author} {\bibinfo {author} {\bibfnamefont {M.}~\bibnamefont {Aspelmeyer}}, \bibinfo {author} {\bibfnamefont {T.~J.}\ \bibnamefont {Kippenberg}},\ and\ \bibinfo {author} {\bibfnamefont {F.}~\bibnamefont {Marquardt}},\ }\href {https://doi.org/10.1103/RevModPhys.86.1391} {\bibfield  {journal} {\bibinfo  {journal} {Reviews of Modern Physics}\ }\textbf {\bibinfo {volume} {86}},\ \bibinfo {pages} {1391} (\bibinfo {year} {2014})}\BibitemShut {NoStop}%
\bibitem [{\citenamefont {Barzanjeh}\ \emph {et~al.}(2022)\citenamefont {Barzanjeh}, \citenamefont {Xuereb}, \citenamefont {Gr{\"o}blacher}, \citenamefont {Paternostro}, \citenamefont {Regal},\ and\ \citenamefont {Weig}}]{barzanjeh2022}%
  \BibitemOpen
  \bibfield  {author} {\bibinfo {author} {\bibfnamefont {S.}~\bibnamefont {Barzanjeh}}, \bibinfo {author} {\bibfnamefont {A.}~\bibnamefont {Xuereb}}, \bibinfo {author} {\bibfnamefont {S.}~\bibnamefont {Gr{\"o}blacher}}, \bibinfo {author} {\bibfnamefont {M.}~\bibnamefont {Paternostro}}, \bibinfo {author} {\bibfnamefont {C.~A.}\ \bibnamefont {Regal}},\ and\ \bibinfo {author} {\bibfnamefont {E.~M.}\ \bibnamefont {Weig}},\ }\href@noop {} {\bibfield  {journal} {\bibinfo  {journal} {Nature Physics}\ }\textbf {\bibinfo {volume} {18}},\ \bibinfo {pages} {15} (\bibinfo {year} {2022})}\BibitemShut {NoStop}%
\bibitem [{\citenamefont {Chan}\ \emph {et~al.}(2011)\citenamefont {Chan}, \citenamefont {Alegre}, \citenamefont {Safavi-Naeini}, \citenamefont {Hill}, \citenamefont {Krause}, \citenamefont {Gröblacher}, \citenamefont {Aspelmeyer},\ and\ \citenamefont {Painter}}]{Chan2011}%
  \BibitemOpen
  \bibfield  {author} {\bibinfo {author} {\bibfnamefont {J.}~\bibnamefont {Chan}}, \bibinfo {author} {\bibfnamefont {T.~P.}\ \bibnamefont {Alegre}}, \bibinfo {author} {\bibfnamefont {A.~H.}\ \bibnamefont {Safavi-Naeini}}, \bibinfo {author} {\bibfnamefont {J.~T.}\ \bibnamefont {Hill}}, \bibinfo {author} {\bibfnamefont {A.}~\bibnamefont {Krause}}, \bibinfo {author} {\bibfnamefont {S.}~\bibnamefont {Gröblacher}}, \bibinfo {author} {\bibfnamefont {M.}~\bibnamefont {Aspelmeyer}},\ and\ \bibinfo {author} {\bibfnamefont {O.}~\bibnamefont {Painter}},\ }\href {https://doi.org/10.1038/nature10461} {\bibfield  {journal} {\bibinfo  {journal} {Nature}\ }\textbf {\bibinfo {volume} {478}},\ \bibinfo {pages} {89} (\bibinfo {year} {2011})}\BibitemShut {NoStop}%
\bibitem [{\citenamefont {Riviere}\ \emph {et~al.}(2011)\citenamefont {Riviere}, \citenamefont {Deleglise}, \citenamefont {Weis}, \citenamefont {Gavartin}, \citenamefont {Arcizet}, \citenamefont {Schliesser},\ and\ \citenamefont {Kippenberg}}]{riviere2011optomechanical}%
  \BibitemOpen
  \bibfield  {author} {\bibinfo {author} {\bibfnamefont {R.}~\bibnamefont {Riviere}}, \bibinfo {author} {\bibfnamefont {S.}~\bibnamefont {Deleglise}}, \bibinfo {author} {\bibfnamefont {S.}~\bibnamefont {Weis}}, \bibinfo {author} {\bibfnamefont {E.}~\bibnamefont {Gavartin}}, \bibinfo {author} {\bibfnamefont {O.}~\bibnamefont {Arcizet}}, \bibinfo {author} {\bibfnamefont {A.}~\bibnamefont {Schliesser}},\ and\ \bibinfo {author} {\bibfnamefont {T.~J.}\ \bibnamefont {Kippenberg}},\ }\href@noop {} {\bibfield  {journal} {\bibinfo  {journal} {Physical Review A---Atomic, Molecular, and Optical Physics}\ }\textbf {\bibinfo {volume} {83}},\ \bibinfo {pages} {063835} (\bibinfo {year} {2011})}\BibitemShut {NoStop}%
\bibitem [{\citenamefont {Li}\ \emph {et~al.}(2021)\citenamefont {Li}, \citenamefont {Ou}, \citenamefont {Lei},\ and\ \citenamefont {Liu}}]{li2021sensing}%
  \BibitemOpen
  \bibfield  {author} {\bibinfo {author} {\bibfnamefont {B.-B.}\ \bibnamefont {Li}}, \bibinfo {author} {\bibfnamefont {L.}~\bibnamefont {Ou}}, \bibinfo {author} {\bibfnamefont {Y.}~\bibnamefont {Lei}},\ and\ \bibinfo {author} {\bibfnamefont {Y.-C.}\ \bibnamefont {Liu}},\ }\href@noop {} {\bibfield  {journal} {\bibinfo  {journal} {Nanophotonics}\ }\textbf {\bibinfo {volume} {10}},\ \bibinfo {pages} {2799} (\bibinfo {year} {2021})}\BibitemShut {NoStop}%
\bibitem [{\citenamefont {Krause}\ \emph {et~al.}(2012)\citenamefont {Krause}, \citenamefont {Winger}, \citenamefont {Blasius}, \citenamefont {Lin},\ and\ \citenamefont {Painter}}]{Krause2012}%
  \BibitemOpen
  \bibfield  {author} {\bibinfo {author} {\bibfnamefont {A.~G.}\ \bibnamefont {Krause}}, \bibinfo {author} {\bibfnamefont {M.}~\bibnamefont {Winger}}, \bibinfo {author} {\bibfnamefont {T.~D.}\ \bibnamefont {Blasius}}, \bibinfo {author} {\bibfnamefont {Q.}~\bibnamefont {Lin}},\ and\ \bibinfo {author} {\bibfnamefont {O.}~\bibnamefont {Painter}},\ }\href@noop {} {\bibfield  {journal} {\bibinfo  {journal} {Nature Photonics}\ }\textbf {\bibinfo {volume} {6}},\ \bibinfo {pages} {768} (\bibinfo {year} {2012})}\BibitemShut {NoStop}%
\bibitem [{\citenamefont {Grudinin}\ \emph {et~al.}(2010)\citenamefont {Grudinin}, \citenamefont {Lee}, \citenamefont {Painter},\ and\ \citenamefont {Vahala}}]{Grudinin2010}%
  \BibitemOpen
  \bibfield  {author} {\bibinfo {author} {\bibfnamefont {I.~S.}\ \bibnamefont {Grudinin}}, \bibinfo {author} {\bibfnamefont {H.}~\bibnamefont {Lee}}, \bibinfo {author} {\bibfnamefont {O.}~\bibnamefont {Painter}},\ and\ \bibinfo {author} {\bibfnamefont {K.~J.}\ \bibnamefont {Vahala}},\ }\href {https://doi.org/10.1103/PhysRevLett.104.083901} {\bibfield  {journal} {\bibinfo  {journal} {Physical Review Letters}\ }\textbf {\bibinfo {volume} {104}},\ \bibinfo {pages} {2} (\bibinfo {year} {2010})}\BibitemShut {NoStop}%
\bibitem [{\citenamefont {Mercad{\'e}}\ \emph {et~al.}(2020)\citenamefont {Mercad{\'e}}, \citenamefont {Mart{\'\i}n}, \citenamefont {Griol}, \citenamefont {Navarro-Urrios},\ and\ \citenamefont {Mart{\'\i}nez}}]{mercade2020microwave}%
  \BibitemOpen
  \bibfield  {author} {\bibinfo {author} {\bibfnamefont {L.}~\bibnamefont {Mercad{\'e}}}, \bibinfo {author} {\bibfnamefont {L.~L.}\ \bibnamefont {Mart{\'\i}n}}, \bibinfo {author} {\bibfnamefont {A.}~\bibnamefont {Griol}}, \bibinfo {author} {\bibfnamefont {D.}~\bibnamefont {Navarro-Urrios}},\ and\ \bibinfo {author} {\bibfnamefont {A.}~\bibnamefont {Mart{\'\i}nez}},\ }\href@noop {} {\bibfield  {journal} {\bibinfo  {journal} {Nanophotonics}\ }\textbf {\bibinfo {volume} {9}},\ \bibinfo {pages} {3535} (\bibinfo {year} {2020})}\BibitemShut {NoStop}%
\bibitem [{\citenamefont {Del’Haye}\ \emph {et~al.}(2007)\citenamefont {Del’Haye}, \citenamefont {Schliesser}, \citenamefont {Arcizet}, \citenamefont {Wilken}, \citenamefont {Holzwarth},\ and\ \citenamefont {Kippenberg}}]{del2007optical}%
  \BibitemOpen
  \bibfield  {author} {\bibinfo {author} {\bibfnamefont {P.}~\bibnamefont {Del’Haye}}, \bibinfo {author} {\bibfnamefont {A.}~\bibnamefont {Schliesser}}, \bibinfo {author} {\bibfnamefont {O.}~\bibnamefont {Arcizet}}, \bibinfo {author} {\bibfnamefont {T.}~\bibnamefont {Wilken}}, \bibinfo {author} {\bibfnamefont {R.}~\bibnamefont {Holzwarth}},\ and\ \bibinfo {author} {\bibfnamefont {T.~J.}\ \bibnamefont {Kippenberg}},\ }\href@noop {} {\bibfield  {journal} {\bibinfo  {journal} {Nature}\ }\textbf {\bibinfo {volume} {450}},\ \bibinfo {pages} {1214} (\bibinfo {year} {2007})}\BibitemShut {NoStop}%
\bibitem [{\citenamefont {Andrews}\ \emph {et~al.}(2014)\citenamefont {Andrews}, \citenamefont {Peterson}, \citenamefont {Purdy}, \citenamefont {Cicak}, \citenamefont {Simmonds}, \citenamefont {Regal},\ and\ \citenamefont {Lehnert}}]{andrews2014bidirectional}%
  \BibitemOpen
  \bibfield  {author} {\bibinfo {author} {\bibfnamefont {R.~W.}\ \bibnamefont {Andrews}}, \bibinfo {author} {\bibfnamefont {R.~W.}\ \bibnamefont {Peterson}}, \bibinfo {author} {\bibfnamefont {T.~P.}\ \bibnamefont {Purdy}}, \bibinfo {author} {\bibfnamefont {K.}~\bibnamefont {Cicak}}, \bibinfo {author} {\bibfnamefont {R.~W.}\ \bibnamefont {Simmonds}}, \bibinfo {author} {\bibfnamefont {C.~A.}\ \bibnamefont {Regal}},\ and\ \bibinfo {author} {\bibfnamefont {K.~W.}\ \bibnamefont {Lehnert}},\ }\href@noop {} {\bibfield  {journal} {\bibinfo  {journal} {Nature physics}\ }\textbf {\bibinfo {volume} {10}},\ \bibinfo {pages} {321} (\bibinfo {year} {2014})}\BibitemShut {NoStop}%
\bibitem [{\citenamefont {Jiang}\ \emph {et~al.}(2020)\citenamefont {Jiang}, \citenamefont {Sarabalis}, \citenamefont {Dahmani}, \citenamefont {Patel}, \citenamefont {Mayor}, \citenamefont {McKenna}, \citenamefont {Van~Laer},\ and\ \citenamefont {Safavi-Naeini}}]{jiang2020efficient}%
  \BibitemOpen
  \bibfield  {author} {\bibinfo {author} {\bibfnamefont {W.}~\bibnamefont {Jiang}}, \bibinfo {author} {\bibfnamefont {C.~J.}\ \bibnamefont {Sarabalis}}, \bibinfo {author} {\bibfnamefont {Y.~D.}\ \bibnamefont {Dahmani}}, \bibinfo {author} {\bibfnamefont {R.~N.}\ \bibnamefont {Patel}}, \bibinfo {author} {\bibfnamefont {F.~M.}\ \bibnamefont {Mayor}}, \bibinfo {author} {\bibfnamefont {T.~P.}\ \bibnamefont {McKenna}}, \bibinfo {author} {\bibfnamefont {R.}~\bibnamefont {Van~Laer}},\ and\ \bibinfo {author} {\bibfnamefont {A.~H.}\ \bibnamefont {Safavi-Naeini}},\ }\href@noop {} {\bibfield  {journal} {\bibinfo  {journal} {Nature communications}\ }\textbf {\bibinfo {volume} {11}},\ \bibinfo {pages} {1166} (\bibinfo {year} {2020})}\BibitemShut {NoStop}%
\bibitem [{\citenamefont {Forsch}\ \emph {et~al.}(2020)\citenamefont {Forsch}, \citenamefont {Stockill}, \citenamefont {Wallucks}, \citenamefont {Marinkovi{\'c}}, \citenamefont {G{\"a}rtner}, \citenamefont {Norte}, \citenamefont {van Otten}, \citenamefont {Fiore}, \citenamefont {Srinivasan},\ and\ \citenamefont {Gr{\"o}blacher}}]{forsch2020microwave}%
  \BibitemOpen
  \bibfield  {author} {\bibinfo {author} {\bibfnamefont {M.}~\bibnamefont {Forsch}}, \bibinfo {author} {\bibfnamefont {R.}~\bibnamefont {Stockill}}, \bibinfo {author} {\bibfnamefont {A.}~\bibnamefont {Wallucks}}, \bibinfo {author} {\bibfnamefont {I.}~\bibnamefont {Marinkovi{\'c}}}, \bibinfo {author} {\bibfnamefont {C.}~\bibnamefont {G{\"a}rtner}}, \bibinfo {author} {\bibfnamefont {R.~A.}\ \bibnamefont {Norte}}, \bibinfo {author} {\bibfnamefont {F.}~\bibnamefont {van Otten}}, \bibinfo {author} {\bibfnamefont {A.}~\bibnamefont {Fiore}}, \bibinfo {author} {\bibfnamefont {K.}~\bibnamefont {Srinivasan}},\ and\ \bibinfo {author} {\bibfnamefont {S.}~\bibnamefont {Gr{\"o}blacher}},\ }\href@noop {} {\bibfield  {journal} {\bibinfo  {journal} {Nature Physics}\ }\textbf {\bibinfo {volume} {16}},\ \bibinfo {pages} {69} (\bibinfo {year} {2020})}\BibitemShut {NoStop}%
\bibitem [{\citenamefont {Han}\ \emph {et~al.}(2021)\citenamefont {Han}, \citenamefont {Fu}, \citenamefont {Zou}, \citenamefont {Jiang},\ and\ \citenamefont {Tang}}]{han2021microwave}%
  \BibitemOpen
  \bibfield  {author} {\bibinfo {author} {\bibfnamefont {X.}~\bibnamefont {Han}}, \bibinfo {author} {\bibfnamefont {W.}~\bibnamefont {Fu}}, \bibinfo {author} {\bibfnamefont {C.-L.}\ \bibnamefont {Zou}}, \bibinfo {author} {\bibfnamefont {L.}~\bibnamefont {Jiang}},\ and\ \bibinfo {author} {\bibfnamefont {H.~X.}\ \bibnamefont {Tang}},\ }\href@noop {} {\bibfield  {journal} {\bibinfo  {journal} {Optica}\ }\textbf {\bibinfo {volume} {8}},\ \bibinfo {pages} {1050} (\bibinfo {year} {2021})}\BibitemShut {NoStop}%
\bibitem [{\citenamefont {Stannigel}\ \emph {et~al.}(2012)\citenamefont {Stannigel}, \citenamefont {Komar}, \citenamefont {Habraken}, \citenamefont {Bennett}, \citenamefont {Lukin}, \citenamefont {Zoller},\ and\ \citenamefont {Rabl}}]{stannigel2012optomechanical}%
  \BibitemOpen
  \bibfield  {author} {\bibinfo {author} {\bibfnamefont {K.}~\bibnamefont {Stannigel}}, \bibinfo {author} {\bibfnamefont {P.}~\bibnamefont {Komar}}, \bibinfo {author} {\bibfnamefont {S.}~\bibnamefont {Habraken}}, \bibinfo {author} {\bibfnamefont {S.}~\bibnamefont {Bennett}}, \bibinfo {author} {\bibfnamefont {M.~D.}\ \bibnamefont {Lukin}}, \bibinfo {author} {\bibfnamefont {P.}~\bibnamefont {Zoller}},\ and\ \bibinfo {author} {\bibfnamefont {P.}~\bibnamefont {Rabl}},\ }\href@noop {} {\bibfield  {journal} {\bibinfo  {journal} {Physical review letters}\ }\textbf {\bibinfo {volume} {109}},\ \bibinfo {pages} {013603} (\bibinfo {year} {2012})}\BibitemShut {NoStop}%
\bibitem [{\citenamefont {Lake}\ \emph {et~al.}(2021)\citenamefont {Lake}, \citenamefont {Mitchell}, \citenamefont {Sukachev},\ and\ \citenamefont {Barclay}}]{lake2021processing}%
  \BibitemOpen
  \bibfield  {author} {\bibinfo {author} {\bibfnamefont {D.~P.}\ \bibnamefont {Lake}}, \bibinfo {author} {\bibfnamefont {M.}~\bibnamefont {Mitchell}}, \bibinfo {author} {\bibfnamefont {D.~D.}\ \bibnamefont {Sukachev}},\ and\ \bibinfo {author} {\bibfnamefont {P.~E.}\ \bibnamefont {Barclay}},\ }\href@noop {} {\bibfield  {journal} {\bibinfo  {journal} {Nature communications}\ }\textbf {\bibinfo {volume} {12}},\ \bibinfo {pages} {663} (\bibinfo {year} {2021})}\BibitemShut {NoStop}%
\bibitem [{\citenamefont {Hill}\ \emph {et~al.}(2012)\citenamefont {Hill}, \citenamefont {Safavi-Naeini}, \citenamefont {Chan},\ and\ \citenamefont {Painter}}]{Hill2012}%
  \BibitemOpen
  \bibfield  {author} {\bibinfo {author} {\bibfnamefont {J.~T.}\ \bibnamefont {Hill}}, \bibinfo {author} {\bibfnamefont {A.~H.}\ \bibnamefont {Safavi-Naeini}}, \bibinfo {author} {\bibfnamefont {J.}~\bibnamefont {Chan}},\ and\ \bibinfo {author} {\bibfnamefont {O.}~\bibnamefont {Painter}},\ }\bibfield  {journal} {\bibinfo  {journal} {Nature Communications}\ }\textbf {\bibinfo {volume} {3}},\ \href {https://doi.org/10.1038/ncomms2201} {10.1038/ncomms2201} (\bibinfo {year} {2012})\BibitemShut {NoStop}%
\bibitem [{\citenamefont {Eichenfield}\ \emph {et~al.}(2009)\citenamefont {Eichenfield}, \citenamefont {Chan}, \citenamefont {Camacho}, \citenamefont {Vahala},\ and\ \citenamefont {Painter}}]{Eichenfield2009}%
  \BibitemOpen
  \bibfield  {author} {\bibinfo {author} {\bibfnamefont {M.}~\bibnamefont {Eichenfield}}, \bibinfo {author} {\bibfnamefont {J.}~\bibnamefont {Chan}}, \bibinfo {author} {\bibfnamefont {R.~M.}\ \bibnamefont {Camacho}}, \bibinfo {author} {\bibfnamefont {K.~J.}\ \bibnamefont {Vahala}},\ and\ \bibinfo {author} {\bibfnamefont {O.}~\bibnamefont {Painter}},\ }\href {https://doi.org/10.1038/nature08524} {\bibfield  {journal} {\bibinfo  {journal} {Nature}\ }\textbf {\bibinfo {volume} {462}},\ \bibinfo {pages} {78} (\bibinfo {year} {2009})}\BibitemShut {NoStop}%
\bibitem [{\citenamefont {Lifshitz}\ and\ \citenamefont {Roukes}(2000)}]{Lifshitz2000TED}%
  \BibitemOpen
  \bibfield  {author} {\bibinfo {author} {\bibfnamefont {R.}~\bibnamefont {Lifshitz}}\ and\ \bibinfo {author} {\bibfnamefont {M.~L.}\ \bibnamefont {Roukes}},\ }\href {https://doi.org/10.1103/PhysRevB.61.5600} {\bibfield  {journal} {\bibinfo  {journal} {Phys. Rev. B}\ }\textbf {\bibinfo {volume} {61}},\ \bibinfo {pages} {5600} (\bibinfo {year} {2000})}\BibitemShut {NoStop}%
\bibitem [{\citenamefont {Ghaffari}\ \emph {et~al.}(2013)\citenamefont {Ghaffari}, \citenamefont {Chandorkar}, \citenamefont {Wang}, \citenamefont {Ng}, \citenamefont {Ahn}, \citenamefont {Hong}, \citenamefont {Yang},\ and\ \citenamefont {Kenny}}]{ghaffari2013quantum}%
  \BibitemOpen
  \bibfield  {author} {\bibinfo {author} {\bibfnamefont {S.}~\bibnamefont {Ghaffari}}, \bibinfo {author} {\bibfnamefont {S.~A.}\ \bibnamefont {Chandorkar}}, \bibinfo {author} {\bibfnamefont {S.}~\bibnamefont {Wang}}, \bibinfo {author} {\bibfnamefont {E.~J.}\ \bibnamefont {Ng}}, \bibinfo {author} {\bibfnamefont {C.~H.}\ \bibnamefont {Ahn}}, \bibinfo {author} {\bibfnamefont {V.}~\bibnamefont {Hong}}, \bibinfo {author} {\bibfnamefont {Y.}~\bibnamefont {Yang}},\ and\ \bibinfo {author} {\bibfnamefont {T.~W.}\ \bibnamefont {Kenny}},\ }\href@noop {} {\bibfield  {journal} {\bibinfo  {journal} {Scientific reports}\ }\textbf {\bibinfo {volume} {3}},\ \bibinfo {pages} {3244} (\bibinfo {year} {2013})}\BibitemShut {NoStop}%
\bibitem [{\citenamefont {Oh}\ \emph {et~al.}(2026)\citenamefont {Oh}, \citenamefont {Dharod}, \citenamefont {Padgett}, \citenamefont {Hughes~Wyatt}, \citenamefont {Venkatraman}, \citenamefont {Parthasarathy}, \citenamefont {Osipova}, \citenamefont {Hedgepeth}, \citenamefont {Cady}, \citenamefont {Basso} \emph {et~al.}}]{oh2026spin}%
  \BibitemOpen
  \bibfield  {author} {\bibinfo {author} {\bibfnamefont {H.}~\bibnamefont {Oh}}, \bibinfo {author} {\bibfnamefont {V.}~\bibnamefont {Dharod}}, \bibinfo {author} {\bibfnamefont {C.}~\bibnamefont {Padgett}}, \bibinfo {author} {\bibfnamefont {L.~B.}\ \bibnamefont {Hughes~Wyatt}}, \bibinfo {author} {\bibfnamefont {J.}~\bibnamefont {Venkatraman}}, \bibinfo {author} {\bibfnamefont {S.}~\bibnamefont {Parthasarathy}}, \bibinfo {author} {\bibfnamefont {E.}~\bibnamefont {Osipova}}, \bibinfo {author} {\bibfnamefont {I.}~\bibnamefont {Hedgepeth}}, \bibinfo {author} {\bibfnamefont {J.~V.}\ \bibnamefont {Cady}}, \bibinfo {author} {\bibfnamefont {L.}~\bibnamefont {Basso}}, \emph {et~al.},\ }\href@noop {} {\bibfield  {journal} {\bibinfo  {journal} {Optica}\ }\textbf {\bibinfo {volume} {13}},\ \bibinfo {pages} {485} (\bibinfo {year} {2026})}\BibitemShut {NoStop}%
\bibitem [{\citenamefont {{Element Six Technologies}}(2024)}]{diamond_handbook}%
  \BibitemOpen
  \bibfield  {author} {\bibinfo {author} {\bibnamefont {{Element Six Technologies}}},\ }\href {https://e6cvd.com/media/wysiwyg/pdf/Diamond_Handbook_2024.pdf} {\emph {\bibinfo {title} {Diamond handbook}}}\ (\bibinfo {year} {2024})\BibitemShut {NoStop}%
\bibitem [{\citenamefont {Kim}\ \emph {et~al.}(2023)\citenamefont {Kim}, \citenamefont {Kurokawa}, \citenamefont {Sakai}, \citenamefont {Koshino}, \citenamefont {Kosaka},\ and\ \citenamefont {Nomura}}]{kim2023diamond}%
  \BibitemOpen
  \bibfield  {author} {\bibinfo {author} {\bibfnamefont {B.}~\bibnamefont {Kim}}, \bibinfo {author} {\bibfnamefont {H.}~\bibnamefont {Kurokawa}}, \bibinfo {author} {\bibfnamefont {K.}~\bibnamefont {Sakai}}, \bibinfo {author} {\bibfnamefont {K.}~\bibnamefont {Koshino}}, \bibinfo {author} {\bibfnamefont {H.}~\bibnamefont {Kosaka}},\ and\ \bibinfo {author} {\bibfnamefont {M.}~\bibnamefont {Nomura}},\ }\href@noop {} {\bibfield  {journal} {\bibinfo  {journal} {Physical Review Applied}\ }\textbf {\bibinfo {volume} {20}},\ \bibinfo {pages} {044037} (\bibinfo {year} {2023})}\BibitemShut {NoStop}%
\bibitem [{\citenamefont {El-Sayed}\ \emph {et~al.}(2026)\citenamefont {El-Sayed}, \citenamefont {Zohari}, \citenamefont {Itoi}, \citenamefont {Parsa}, \citenamefont {Luiz}, \citenamefont {Losby}, \citenamefont {Hayashida}, \citenamefont {Malac},\ and\ \citenamefont {Barclay}}]{elsayed2026exceptional}%
  \BibitemOpen
  \bibfield  {author} {\bibinfo {author} {\bibfnamefont {W.}~\bibnamefont {El-Sayed}}, \bibinfo {author} {\bibfnamefont {E.}~\bibnamefont {Zohari}}, \bibinfo {author} {\bibfnamefont {J.}~\bibnamefont {Itoi}}, \bibinfo {author} {\bibfnamefont {P.}~\bibnamefont {Parsa}}, \bibinfo {author} {\bibfnamefont {G.~d.~O.}\ \bibnamefont {Luiz}}, \bibinfo {author} {\bibfnamefont {J.~E.}\ \bibnamefont {Losby}}, \bibinfo {author} {\bibfnamefont {M.}~\bibnamefont {Hayashida}}, \bibinfo {author} {\bibfnamefont {M.}~\bibnamefont {Malac}},\ and\ \bibinfo {author} {\bibfnamefont {P.~E.}\ \bibnamefont {Barclay}},\ }\href@noop {} {\bibfield  {journal} {\bibinfo  {journal} {arXiv preprint arXiv:2605.27536}\ } (\bibinfo {year} {2026})}\BibitemShut {NoStop}%
\bibitem [{\citenamefont {Awschalom}\ \emph {et~al.}(2018)\citenamefont {Awschalom}, \citenamefont {Hanson}, \citenamefont {Wrachtrup},\ and\ \citenamefont {Zhou}}]{awschalom2018quantum}%
  \BibitemOpen
  \bibfield  {author} {\bibinfo {author} {\bibfnamefont {D.~D.}\ \bibnamefont {Awschalom}}, \bibinfo {author} {\bibfnamefont {R.}~\bibnamefont {Hanson}}, \bibinfo {author} {\bibfnamefont {J.}~\bibnamefont {Wrachtrup}},\ and\ \bibinfo {author} {\bibfnamefont {B.~B.}\ \bibnamefont {Zhou}},\ }\href@noop {} {\bibfield  {journal} {\bibinfo  {journal} {Nature Photonics}\ }\textbf {\bibinfo {volume} {12}},\ \bibinfo {pages} {516} (\bibinfo {year} {2018})}\BibitemShut {NoStop}%
\bibitem [{\citenamefont {Doherty}\ \emph {et~al.}(2013)\citenamefont {Doherty}, \citenamefont {Manson}, \citenamefont {Delaney}, \citenamefont {Jelezko}, \citenamefont {Wrachtrup},\ and\ \citenamefont {Hollenberg}}]{doherty2013nitrogen}%
  \BibitemOpen
  \bibfield  {author} {\bibinfo {author} {\bibfnamefont {M.~W.}\ \bibnamefont {Doherty}}, \bibinfo {author} {\bibfnamefont {N.~B.}\ \bibnamefont {Manson}}, \bibinfo {author} {\bibfnamefont {P.}~\bibnamefont {Delaney}}, \bibinfo {author} {\bibfnamefont {F.}~\bibnamefont {Jelezko}}, \bibinfo {author} {\bibfnamefont {J.}~\bibnamefont {Wrachtrup}},\ and\ \bibinfo {author} {\bibfnamefont {L.~C.}\ \bibnamefont {Hollenberg}},\ }\href@noop {} {\bibfield  {journal} {\bibinfo  {journal} {Physics Reports}\ }\textbf {\bibinfo {volume} {528}},\ \bibinfo {pages} {1} (\bibinfo {year} {2013})}\BibitemShut {NoStop}%
\bibitem [{\citenamefont {MacQuarrie}\ \emph {et~al.}(2013)\citenamefont {MacQuarrie}, \citenamefont {Gosavi}, \citenamefont {Jungwirth}, \citenamefont {Bhave},\ and\ \citenamefont {Fuchs}}]{macquarrie2013}%
  \BibitemOpen
  \bibfield  {author} {\bibinfo {author} {\bibfnamefont {E.}~\bibnamefont {MacQuarrie}}, \bibinfo {author} {\bibfnamefont {T.}~\bibnamefont {Gosavi}}, \bibinfo {author} {\bibfnamefont {N.}~\bibnamefont {Jungwirth}}, \bibinfo {author} {\bibfnamefont {S.}~\bibnamefont {Bhave}},\ and\ \bibinfo {author} {\bibfnamefont {G.}~\bibnamefont {Fuchs}},\ }\href@noop {} {\bibfield  {journal} {\bibinfo  {journal} {Physical review letters}\ }\textbf {\bibinfo {volume} {111}},\ \bibinfo {pages} {227602} (\bibinfo {year} {2013})}\BibitemShut {NoStop}%
\bibitem [{\citenamefont {Shandilya}\ \emph {et~al.}(2021)\citenamefont {Shandilya}, \citenamefont {Lake}, \citenamefont {Mitchell}, \citenamefont {Sukachev},\ and\ \citenamefont {Barclay}}]{shandilya2021optomechanical}%
  \BibitemOpen
  \bibfield  {author} {\bibinfo {author} {\bibfnamefont {P.~K.}\ \bibnamefont {Shandilya}}, \bibinfo {author} {\bibfnamefont {D.~P.}\ \bibnamefont {Lake}}, \bibinfo {author} {\bibfnamefont {M.~J.}\ \bibnamefont {Mitchell}}, \bibinfo {author} {\bibfnamefont {D.~D.}\ \bibnamefont {Sukachev}},\ and\ \bibinfo {author} {\bibfnamefont {P.~E.}\ \bibnamefont {Barclay}},\ }\href@noop {} {\bibfield  {journal} {\bibinfo  {journal} {Nature Physics}\ }\textbf {\bibinfo {volume} {17}},\ \bibinfo {pages} {1420} (\bibinfo {year} {2021})}\BibitemShut {NoStop}%
\bibitem [{\citenamefont {Ovartchaiyapong}\ \emph {et~al.}(2014)\citenamefont {Ovartchaiyapong}, \citenamefont {Lee}, \citenamefont {Myers},\ and\ \citenamefont {Jayich}}]{ovartchaiyapong2014dynamic}%
  \BibitemOpen
  \bibfield  {author} {\bibinfo {author} {\bibfnamefont {P.}~\bibnamefont {Ovartchaiyapong}}, \bibinfo {author} {\bibfnamefont {K.~W.}\ \bibnamefont {Lee}}, \bibinfo {author} {\bibfnamefont {B.~A.}\ \bibnamefont {Myers}},\ and\ \bibinfo {author} {\bibfnamefont {A.~C.~B.}\ \bibnamefont {Jayich}},\ }\href@noop {} {\bibfield  {journal} {\bibinfo  {journal} {Nature communications}\ }\textbf {\bibinfo {volume} {5}},\ \bibinfo {pages} {4429} (\bibinfo {year} {2014})}\BibitemShut {NoStop}%
\bibitem [{\citenamefont {Udvarhelyi}\ \emph {et~al.}(2018)\citenamefont {Udvarhelyi}, \citenamefont {Shkolnikov}, \citenamefont {Gali}, \citenamefont {Burkard},\ and\ \citenamefont {P\'alyi}}]{udvarhelyi2017spin}%
  \BibitemOpen
  \bibfield  {author} {\bibinfo {author} {\bibfnamefont {P.}~\bibnamefont {Udvarhelyi}}, \bibinfo {author} {\bibfnamefont {V.~O.}\ \bibnamefont {Shkolnikov}}, \bibinfo {author} {\bibfnamefont {A.}~\bibnamefont {Gali}}, \bibinfo {author} {\bibfnamefont {G.}~\bibnamefont {Burkard}},\ and\ \bibinfo {author} {\bibfnamefont {A.}~\bibnamefont {P\'alyi}},\ }\href {https://doi.org/10.1103/PhysRevB.98.075201} {\bibfield  {journal} {\bibinfo  {journal} {Phys. Rev. B}\ }\textbf {\bibinfo {volume} {98}},\ \bibinfo {pages} {075201} (\bibinfo {year} {2018})}\BibitemShut {NoStop}%
\bibitem [{\citenamefont {Teissier}\ \emph {et~al.}(2014)\citenamefont {Teissier}, \citenamefont {Barfuss}, \citenamefont {Appel}, \citenamefont {Neu},\ and\ \citenamefont {Maletinsky}}]{teissier2014strain}%
  \BibitemOpen
  \bibfield  {author} {\bibinfo {author} {\bibfnamefont {J.}~\bibnamefont {Teissier}}, \bibinfo {author} {\bibfnamefont {A.}~\bibnamefont {Barfuss}}, \bibinfo {author} {\bibfnamefont {P.}~\bibnamefont {Appel}}, \bibinfo {author} {\bibfnamefont {E.}~\bibnamefont {Neu}},\ and\ \bibinfo {author} {\bibfnamefont {P.}~\bibnamefont {Maletinsky}},\ }\href@noop {} {\bibfield  {journal} {\bibinfo  {journal} {Physical review letters}\ }\textbf {\bibinfo {volume} {113}},\ \bibinfo {pages} {020503} (\bibinfo {year} {2014})}\BibitemShut {NoStop}%
\bibitem [{\citenamefont {Raniwala}\ \emph {et~al.}(2025)\citenamefont {Raniwala}, \citenamefont {Anand}, \citenamefont {Krastanov}, \citenamefont {Eichenfield}, \citenamefont {Trusheim},\ and\ \citenamefont {Englund}}]{raniwala2025spin}%
  \BibitemOpen
  \bibfield  {author} {\bibinfo {author} {\bibfnamefont {H.}~\bibnamefont {Raniwala}}, \bibinfo {author} {\bibfnamefont {P.}~\bibnamefont {Anand}}, \bibinfo {author} {\bibfnamefont {S.}~\bibnamefont {Krastanov}}, \bibinfo {author} {\bibfnamefont {M.}~\bibnamefont {Eichenfield}}, \bibinfo {author} {\bibfnamefont {M.}~\bibnamefont {Trusheim}},\ and\ \bibinfo {author} {\bibfnamefont {D.~R.}\ \bibnamefont {Englund}},\ }\href@noop {} {\bibfield  {journal} {\bibinfo  {journal} {npj Quantum Information}\ }\textbf {\bibinfo {volume} {11}},\ \bibinfo {pages} {120} (\bibinfo {year} {2025})}\BibitemShut {NoStop}%
\bibitem [{\citenamefont {Joe}\ \emph {et~al.}(2026)\citenamefont {Joe}, \citenamefont {Haas}, \citenamefont {Kuruma}, \citenamefont {Jin}, \citenamefont {Kang}, \citenamefont {Ding}, \citenamefont {Chia}, \citenamefont {Warner}, \citenamefont {Pingault}, \citenamefont {Machielse} \emph {et~al.}}]{joe2026purcell}%
  \BibitemOpen
  \bibfield  {author} {\bibinfo {author} {\bibfnamefont {G.}~\bibnamefont {Joe}}, \bibinfo {author} {\bibfnamefont {M.}~\bibnamefont {Haas}}, \bibinfo {author} {\bibfnamefont {K.}~\bibnamefont {Kuruma}}, \bibinfo {author} {\bibfnamefont {C.}~\bibnamefont {Jin}}, \bibinfo {author} {\bibfnamefont {D.~D.}\ \bibnamefont {Kang}}, \bibinfo {author} {\bibfnamefont {S.~W.}\ \bibnamefont {Ding}}, \bibinfo {author} {\bibfnamefont {C.}~\bibnamefont {Chia}}, \bibinfo {author} {\bibfnamefont {H.}~\bibnamefont {Warner}}, \bibinfo {author} {\bibfnamefont {B.}~\bibnamefont {Pingault}}, \bibinfo {author} {\bibfnamefont {B.}~\bibnamefont {Machielse}}, \emph {et~al.},\ }\href@noop {} {\bibfield  {journal} {\bibinfo  {journal} {Nature}\ }\textbf {\bibinfo {volume} {653}},\ \bibinfo {pages} {378} (\bibinfo {year} {2026})}\BibitemShut {NoStop}%
\bibitem [{\citenamefont {Shaw}\ \emph {et~al.}(1994)\citenamefont {Shaw}, \citenamefont {Zhang},\ and\ \citenamefont {MacDonald}}]{SHAW1994}%
  \BibitemOpen
  \bibfield  {author} {\bibinfo {author} {\bibfnamefont {K.~A.}\ \bibnamefont {Shaw}}, \bibinfo {author} {\bibfnamefont {Z.}~\bibnamefont {Zhang}},\ and\ \bibinfo {author} {\bibfnamefont {N.~C.}\ \bibnamefont {MacDonald}},\ }\href {https://doi.org/https://doi.org/10.1016/0924-4247(94)85031-3} {\bibfield  {journal} {\bibinfo  {journal} {Sensors and Actuators A: Physical}\ }\textbf {\bibinfo {volume} {40}},\ \bibinfo {pages} {63} (\bibinfo {year} {1994})}\BibitemShut {NoStop}%
\bibitem [{\citenamefont {Khanaliloo}\ \emph {et~al.}(2015{\natexlab{a}})\citenamefont {Khanaliloo}, \citenamefont {Mitchell}, \citenamefont {Hryciw},\ and\ \citenamefont {Barclay}}]{Khanaliloo2015MD}%
  \BibitemOpen
  \bibfield  {author} {\bibinfo {author} {\bibfnamefont {B.}~\bibnamefont {Khanaliloo}}, \bibinfo {author} {\bibfnamefont {M.}~\bibnamefont {Mitchell}}, \bibinfo {author} {\bibfnamefont {A.~C.}\ \bibnamefont {Hryciw}},\ and\ \bibinfo {author} {\bibfnamefont {P.~E.}\ \bibnamefont {Barclay}},\ }\href {https://doi.org/10.1021/acs.nanolett.5b01346} {\bibfield  {journal} {\bibinfo  {journal} {Nano Letters}\ }\textbf {\bibinfo {volume} {15}},\ \bibinfo {pages} {5131} (\bibinfo {year} {2015}{\natexlab{a}})},\ \Eprint {https://arxiv.org/abs/https://doi.org/10.1021/acs.nanolett.5b01346} {https://doi.org/10.1021/acs.nanolett.5b01346} \BibitemShut {NoStop}%
\bibitem [{\citenamefont {Khanaliloo}\ \emph {et~al.}(2015{\natexlab{b}})\citenamefont {Khanaliloo}, \citenamefont {Jayakumar}, \citenamefont {Hryciw}, \citenamefont {Lake}, \citenamefont {Kaviani},\ and\ \citenamefont {Barclay}}]{Khanaliloo2015NB}%
  \BibitemOpen
  \bibfield  {author} {\bibinfo {author} {\bibfnamefont {B.}~\bibnamefont {Khanaliloo}}, \bibinfo {author} {\bibfnamefont {H.}~\bibnamefont {Jayakumar}}, \bibinfo {author} {\bibfnamefont {A.~C.}\ \bibnamefont {Hryciw}}, \bibinfo {author} {\bibfnamefont {D.~P.}\ \bibnamefont {Lake}}, \bibinfo {author} {\bibfnamefont {H.}~\bibnamefont {Kaviani}},\ and\ \bibinfo {author} {\bibfnamefont {P.~E.}\ \bibnamefont {Barclay}},\ }\href {https://login.ezproxy.library.ualberta.ca/login?url=https://www.proquest.com/scholarly-journals/single-crystal-diamond-nanobeam-waveguide/docview/2550553456/se-2} {\bibfield  {journal} {\bibinfo  {journal} {Physical Review.X}\ }\textbf {\bibinfo {volume} {5}} (\bibinfo {year} {2015}{\natexlab{b}})}\BibitemShut {NoStop}%
\bibitem [{\citenamefont {Mouradian}\ \emph {et~al.}(2017)\citenamefont {Mouradian}, \citenamefont {Wan}, \citenamefont {Schröder},\ and\ \citenamefont {Englund}}]{Mouradian2017}%
  \BibitemOpen
  \bibfield  {author} {\bibinfo {author} {\bibfnamefont {S.}~\bibnamefont {Mouradian}}, \bibinfo {author} {\bibfnamefont {N.~H.}\ \bibnamefont {Wan}}, \bibinfo {author} {\bibfnamefont {T.}~\bibnamefont {Schröder}},\ and\ \bibinfo {author} {\bibfnamefont {D.}~\bibnamefont {Englund}},\ }\bibfield  {journal} {\bibinfo  {journal} {Applied Physics Letters}\ }\textbf {\bibinfo {volume} {111}},\ \href {https://doi.org/10.1063/1.4992118} {10.1063/1.4992118} (\bibinfo {year} {2017})\BibitemShut {NoStop}%
\bibitem [{\citenamefont {Wan}\ \emph {et~al.}(2018)\citenamefont {Wan}, \citenamefont {Mouradian},\ and\ \citenamefont {Englund}}]{wan2018two}%
  \BibitemOpen
  \bibfield  {author} {\bibinfo {author} {\bibfnamefont {N.~H.}\ \bibnamefont {Wan}}, \bibinfo {author} {\bibfnamefont {S.}~\bibnamefont {Mouradian}},\ and\ \bibinfo {author} {\bibfnamefont {D.}~\bibnamefont {Englund}},\ }\href@noop {} {\bibfield  {journal} {\bibinfo  {journal} {Applied Physics Letters}\ }\textbf {\bibinfo {volume} {112}} (\bibinfo {year} {2018})}\BibitemShut {NoStop}%
\bibitem [{\citenamefont {Safavi-Naeini}\ and\ \citenamefont {Painter}(2010)}]{Safavi2010}%
  \BibitemOpen
  \bibfield  {author} {\bibinfo {author} {\bibfnamefont {A.~H.}\ \bibnamefont {Safavi-Naeini}}\ and\ \bibinfo {author} {\bibfnamefont {O.}~\bibnamefont {Painter}},\ }\href@noop {} {\bibfield  {journal} {\bibinfo  {journal} {Optics express}\ }\textbf {\bibinfo {volume} {18}},\ \bibinfo {pages} {14926} (\bibinfo {year} {2010})}\BibitemShut {NoStop}%
\bibitem [{\citenamefont {Chan}\ \emph {et~al.}(2012)\citenamefont {Chan}, \citenamefont {Safavi-Naeini}, \citenamefont {Hill}, \citenamefont {Meenehan},\ and\ \citenamefont {Painter}}]{Chan2012}%
  \BibitemOpen
  \bibfield  {author} {\bibinfo {author} {\bibfnamefont {J.}~\bibnamefont {Chan}}, \bibinfo {author} {\bibfnamefont {A.~H.}\ \bibnamefont {Safavi-Naeini}}, \bibinfo {author} {\bibfnamefont {J.~T.}\ \bibnamefont {Hill}}, \bibinfo {author} {\bibfnamefont {S.}~\bibnamefont {Meenehan}},\ and\ \bibinfo {author} {\bibfnamefont {O.}~\bibnamefont {Painter}},\ }\bibfield  {journal} {\bibinfo  {journal} {Applied Physics Letters}\ }\textbf {\bibinfo {volume} {101}},\ \href {https://doi.org/10.1063/1.4747726} {10.1063/1.4747726} (\bibinfo {year} {2012})\BibitemShut {NoStop}%
\bibitem [{\citenamefont {Moraes}\ \emph {et~al.}(2022)\citenamefont {Moraes}, \citenamefont {de~Aguiar}, \citenamefont {De~Melo}, \citenamefont {Wiederhecker},\ and\ \citenamefont {Alegre}}]{moraes2022optimization}%
  \BibitemOpen
  \bibfield  {author} {\bibinfo {author} {\bibfnamefont {F.}~\bibnamefont {Moraes}}, \bibinfo {author} {\bibfnamefont {G.~H.}\ \bibnamefont {de~Aguiar}}, \bibinfo {author} {\bibfnamefont {E.~G.}\ \bibnamefont {De~Melo}}, \bibinfo {author} {\bibfnamefont {G.~S.}\ \bibnamefont {Wiederhecker}},\ and\ \bibinfo {author} {\bibfnamefont {T.~P.~M.}\ \bibnamefont {Alegre}},\ }\href@noop {} {\bibfield  {journal} {\bibinfo  {journal} {Journal of the Optical Society of America B}\ }\textbf {\bibinfo {volume} {39}},\ \bibinfo {pages} {2735} (\bibinfo {year} {2022})}\BibitemShut {NoStop}%
\bibitem [{\citenamefont {Quan}\ and\ \citenamefont {Loncar}(2011)}]{quan2011}%
  \BibitemOpen
  \bibfield  {author} {\bibinfo {author} {\bibfnamefont {Q.}~\bibnamefont {Quan}}\ and\ \bibinfo {author} {\bibfnamefont {M.}~\bibnamefont {Loncar}},\ }\href@noop {} {\bibfield  {journal} {\bibinfo  {journal} {Optics express}\ }\textbf {\bibinfo {volume} {19}},\ \bibinfo {pages} {18529} (\bibinfo {year} {2011})}\BibitemShut {NoStop}%
\bibitem [{\citenamefont {Burek}\ \emph {et~al.}(2016)\citenamefont {Burek}, \citenamefont {Cohen}, \citenamefont {Meenehan}, \citenamefont {El-Sawah}, \citenamefont {Chia}, \citenamefont {Ruelle}, \citenamefont {Meesala}, \citenamefont {Rochman}, \citenamefont {Atikian}, \citenamefont {Markham}, \citenamefont {Twitchen}, \citenamefont {Lukin}, \citenamefont {Painter},\ and\ \citenamefont {Lončar}}]{Burek2016}%
  \BibitemOpen
  \bibfield  {author} {\bibinfo {author} {\bibfnamefont {M.~J.}\ \bibnamefont {Burek}}, \bibinfo {author} {\bibfnamefont {J.~D.}\ \bibnamefont {Cohen}}, \bibinfo {author} {\bibfnamefont {S.~M.}\ \bibnamefont {Meenehan}}, \bibinfo {author} {\bibfnamefont {N.}~\bibnamefont {El-Sawah}}, \bibinfo {author} {\bibfnamefont {C.}~\bibnamefont {Chia}}, \bibinfo {author} {\bibfnamefont {T.}~\bibnamefont {Ruelle}}, \bibinfo {author} {\bibfnamefont {S.}~\bibnamefont {Meesala}}, \bibinfo {author} {\bibfnamefont {J.}~\bibnamefont {Rochman}}, \bibinfo {author} {\bibfnamefont {H.~A.}\ \bibnamefont {Atikian}}, \bibinfo {author} {\bibfnamefont {M.}~\bibnamefont {Markham}}, \bibinfo {author} {\bibfnamefont {D.~J.}\ \bibnamefont {Twitchen}}, \bibinfo {author} {\bibfnamefont {M.~D.}\ \bibnamefont {Lukin}}, \bibinfo {author} {\bibfnamefont {O.}~\bibnamefont {Painter}},\ and\ \bibinfo {author} {\bibfnamefont {M.}~\bibnamefont {Lončar}},\ }\href {https://doi.org/10.1364/optica.3.001404} {\bibfield  {journal} {\bibinfo  {journal}
  {Optica}\ }\textbf {\bibinfo {volume} {3}},\ \bibinfo {pages} {1404} (\bibinfo {year} {2016})}\BibitemShut {NoStop}%
\bibitem [{\citenamefont {Michael}\ \emph {et~al.}(2007)\citenamefont {Michael}, \citenamefont {Borselli}, \citenamefont {Johnson}, \citenamefont {Chrystal},\ and\ \citenamefont {Painter}}]{michael2007}%
  \BibitemOpen
  \bibfield  {author} {\bibinfo {author} {\bibfnamefont {C.~P.}\ \bibnamefont {Michael}}, \bibinfo {author} {\bibfnamefont {M.}~\bibnamefont {Borselli}}, \bibinfo {author} {\bibfnamefont {T.~J.}\ \bibnamefont {Johnson}}, \bibinfo {author} {\bibfnamefont {C.}~\bibnamefont {Chrystal}},\ and\ \bibinfo {author} {\bibfnamefont {O.}~\bibnamefont {Painter}},\ }\href@noop {} {\bibfield  {journal} {\bibinfo  {journal} {Optics express}\ }\textbf {\bibinfo {volume} {15}},\ \bibinfo {pages} {4745} (\bibinfo {year} {2007})}\BibitemShut {NoStop}%
\bibitem [{\citenamefont {Marquardt}\ \emph {et~al.}(2007)\citenamefont {Marquardt}, \citenamefont {Chen}, \citenamefont {Clerk},\ and\ \citenamefont {Girvin}}]{marquardt2007}%
  \BibitemOpen
  \bibfield  {author} {\bibinfo {author} {\bibfnamefont {F.}~\bibnamefont {Marquardt}}, \bibinfo {author} {\bibfnamefont {J.~P.}\ \bibnamefont {Chen}}, \bibinfo {author} {\bibfnamefont {A.~A.}\ \bibnamefont {Clerk}},\ and\ \bibinfo {author} {\bibfnamefont {S.}~\bibnamefont {Girvin}},\ }\href@noop {} {\bibfield  {journal} {\bibinfo  {journal} {Physical review letters}\ }\textbf {\bibinfo {volume} {99}},\ \bibinfo {pages} {093902} (\bibinfo {year} {2007})}\BibitemShut {NoStop}%
\bibitem [{\citenamefont {Joe}\ \emph {et~al.}(2024)\citenamefont {Joe}, \citenamefont {Chia}, \citenamefont {Pingault}, \citenamefont {Haas}, \citenamefont {Chalupnik}, \citenamefont {Cornell}, \citenamefont {Kuruma}, \citenamefont {Machielse}, \citenamefont {Sinclair}, \citenamefont {Meesala},\ and\ \citenamefont {Lončar}}]{Joe2024}%
  \BibitemOpen
  \bibfield  {author} {\bibinfo {author} {\bibfnamefont {G.}~\bibnamefont {Joe}}, \bibinfo {author} {\bibfnamefont {C.}~\bibnamefont {Chia}}, \bibinfo {author} {\bibfnamefont {B.}~\bibnamefont {Pingault}}, \bibinfo {author} {\bibfnamefont {M.}~\bibnamefont {Haas}}, \bibinfo {author} {\bibfnamefont {M.}~\bibnamefont {Chalupnik}}, \bibinfo {author} {\bibfnamefont {E.}~\bibnamefont {Cornell}}, \bibinfo {author} {\bibfnamefont {K.}~\bibnamefont {Kuruma}}, \bibinfo {author} {\bibfnamefont {B.}~\bibnamefont {Machielse}}, \bibinfo {author} {\bibfnamefont {N.}~\bibnamefont {Sinclair}}, \bibinfo {author} {\bibfnamefont {S.}~\bibnamefont {Meesala}},\ and\ \bibinfo {author} {\bibfnamefont {M.}~\bibnamefont {Lončar}},\ }\href {https://doi.org/10.1021/acs.nanolett.3c04953} {\bibfield  {journal} {\bibinfo  {journal} {Nano Letters}\ }\textbf {\bibinfo {volume} {24}},\ \bibinfo {pages} {6831} (\bibinfo {year} {2024})}\BibitemShut {NoStop}%
\bibitem [{\citenamefont {Cady}\ \emph {et~al.}(2019)\citenamefont {Cady}, \citenamefont {Michel}, \citenamefont {Lee}, \citenamefont {Patel}, \citenamefont {Sarabalis}, \citenamefont {Safavi-Naeini},\ and\ \citenamefont {Jayich}}]{cady2019diamond}%
  \BibitemOpen
  \bibfield  {author} {\bibinfo {author} {\bibfnamefont {J.~V.}\ \bibnamefont {Cady}}, \bibinfo {author} {\bibfnamefont {O.}~\bibnamefont {Michel}}, \bibinfo {author} {\bibfnamefont {K.~W.}\ \bibnamefont {Lee}}, \bibinfo {author} {\bibfnamefont {R.~N.}\ \bibnamefont {Patel}}, \bibinfo {author} {\bibfnamefont {C.~J.}\ \bibnamefont {Sarabalis}}, \bibinfo {author} {\bibfnamefont {A.~H.}\ \bibnamefont {Safavi-Naeini}},\ and\ \bibinfo {author} {\bibfnamefont {A.~C.~B.}\ \bibnamefont {Jayich}},\ }\href@noop {} {\bibfield  {journal} {\bibinfo  {journal} {Quantum Science and Technology}\ }\textbf {\bibinfo {volume} {4}},\ \bibinfo {pages} {024009} (\bibinfo {year} {2019})}\BibitemShut {NoStop}%
\bibitem [{\citenamefont {Norte}\ \emph {et~al.}(2016)\citenamefont {Norte}, \citenamefont {Moura},\ and\ \citenamefont {Gr{\"o}blacher}}]{norte2016mechanical}%
  \BibitemOpen
  \bibfield  {author} {\bibinfo {author} {\bibfnamefont {R.~A.}\ \bibnamefont {Norte}}, \bibinfo {author} {\bibfnamefont {J.~P.}\ \bibnamefont {Moura}},\ and\ \bibinfo {author} {\bibfnamefont {S.}~\bibnamefont {Gr{\"o}blacher}},\ }\href@noop {} {\bibfield  {journal} {\bibinfo  {journal} {Physical review letters}\ }\textbf {\bibinfo {volume} {116}},\ \bibinfo {pages} {147202} (\bibinfo {year} {2016})}\BibitemShut {NoStop}%
\bibitem [{\citenamefont {Khosla}\ \emph {et~al.}(2017)\citenamefont {Khosla}, \citenamefont {Brawley}, \citenamefont {Vanner},\ and\ \citenamefont {Bowen}}]{khosla2017quantum}%
  \BibitemOpen
  \bibfield  {author} {\bibinfo {author} {\bibfnamefont {K.~E.}\ \bibnamefont {Khosla}}, \bibinfo {author} {\bibfnamefont {G.~A.}\ \bibnamefont {Brawley}}, \bibinfo {author} {\bibfnamefont {M.~R.}\ \bibnamefont {Vanner}},\ and\ \bibinfo {author} {\bibfnamefont {W.~P.}\ \bibnamefont {Bowen}},\ }\href@noop {} {\bibfield  {journal} {\bibinfo  {journal} {Optica}\ }\textbf {\bibinfo {volume} {4}},\ \bibinfo {pages} {1382} (\bibinfo {year} {2017})}\BibitemShut {NoStop}%
\bibitem [{\citenamefont {Tsaturyan}\ \emph {et~al.}(2017)\citenamefont {Tsaturyan}, \citenamefont {Barg}, \citenamefont {Polzik},\ and\ \citenamefont {Schliesser}}]{tsaturyan2017ultracoherent}%
  \BibitemOpen
  \bibfield  {author} {\bibinfo {author} {\bibfnamefont {Y.}~\bibnamefont {Tsaturyan}}, \bibinfo {author} {\bibfnamefont {A.}~\bibnamefont {Barg}}, \bibinfo {author} {\bibfnamefont {E.~S.}\ \bibnamefont {Polzik}},\ and\ \bibinfo {author} {\bibfnamefont {A.}~\bibnamefont {Schliesser}},\ }\href@noop {} {\bibfield  {journal} {\bibinfo  {journal} {Nature nanotechnology}\ }\textbf {\bibinfo {volume} {12}},\ \bibinfo {pages} {776} (\bibinfo {year} {2017})}\BibitemShut {NoStop}%
\bibitem [{\citenamefont {Ren}\ \emph {et~al.}(2020)\citenamefont {Ren}, \citenamefont {Matheny}, \citenamefont {MacCabe}, \citenamefont {Luo}, \citenamefont {Pfeifer}, \citenamefont {Mirhosseini},\ and\ \citenamefont {Painter}}]{Ren2020}%
  \BibitemOpen
  \bibfield  {author} {\bibinfo {author} {\bibfnamefont {H.}~\bibnamefont {Ren}}, \bibinfo {author} {\bibfnamefont {M.~H.}\ \bibnamefont {Matheny}}, \bibinfo {author} {\bibfnamefont {G.~S.}\ \bibnamefont {MacCabe}}, \bibinfo {author} {\bibfnamefont {J.}~\bibnamefont {Luo}}, \bibinfo {author} {\bibfnamefont {H.}~\bibnamefont {Pfeifer}}, \bibinfo {author} {\bibfnamefont {M.}~\bibnamefont {Mirhosseini}},\ and\ \bibinfo {author} {\bibfnamefont {O.}~\bibnamefont {Painter}},\ }\bibfield  {journal} {\bibinfo  {journal} {Nature Communications}\ }\textbf {\bibinfo {volume} {11}},\ \href {https://doi.org/10.1038/s41467-020-17182-9} {10.1038/s41467-020-17182-9} (\bibinfo {year} {2020})\BibitemShut {NoStop}%
\bibitem [{\citenamefont {Marquardt}\ \emph {et~al.}(2006)\citenamefont {Marquardt}, \citenamefont {Harris},\ and\ \citenamefont {Girvin}}]{Marquardt2006}%
  \BibitemOpen
  \bibfield  {author} {\bibinfo {author} {\bibfnamefont {F.}~\bibnamefont {Marquardt}}, \bibinfo {author} {\bibfnamefont {J.~G.}\ \bibnamefont {Harris}},\ and\ \bibinfo {author} {\bibfnamefont {S.~M.}\ \bibnamefont {Girvin}},\ }\bibfield  {journal} {\bibinfo  {journal} {Physical Review Letters}\ }\textbf {\bibinfo {volume} {96}},\ \href {https://doi.org/10.1103/PhysRevLett.96.103901} {10.1103/PhysRevLett.96.103901} (\bibinfo {year} {2006})\BibitemShut {NoStop}%
\bibitem [{\citenamefont {Kippenberg}\ and\ \citenamefont {Vahala}(2008)}]{Kippenberg2008}%
  \BibitemOpen
  \bibfield  {author} {\bibinfo {author} {\bibfnamefont {T.~J.}\ \bibnamefont {Kippenberg}}\ and\ \bibinfo {author} {\bibfnamefont {K.~J.}\ \bibnamefont {Vahala}},\ }\href@noop {} {\bibfield  {journal} {\bibinfo  {journal} {science}\ }\textbf {\bibinfo {volume} {321}},\ \bibinfo {pages} {1172} (\bibinfo {year} {2008})}\BibitemShut {NoStop}%
\bibitem [{\citenamefont {Carmon}\ \emph {et~al.}(2005)\citenamefont {Carmon}, \citenamefont {Rokhsari}, \citenamefont {Yang}, \citenamefont {Kippenberg},\ and\ \citenamefont {Vahala}}]{Carmon2005}%
  \BibitemOpen
  \bibfield  {author} {\bibinfo {author} {\bibfnamefont {T.}~\bibnamefont {Carmon}}, \bibinfo {author} {\bibfnamefont {H.}~\bibnamefont {Rokhsari}}, \bibinfo {author} {\bibfnamefont {L.}~\bibnamefont {Yang}}, \bibinfo {author} {\bibfnamefont {T.~J.}\ \bibnamefont {Kippenberg}},\ and\ \bibinfo {author} {\bibfnamefont {K.~J.}\ \bibnamefont {Vahala}},\ }\bibfield  {journal} {\bibinfo  {journal} {Physical Review Letters}\ }\textbf {\bibinfo {volume} {94}},\ \href {https://doi.org/10.1103/PhysRevLett.94.223902} {10.1103/PhysRevLett.94.223902} (\bibinfo {year} {2005})\BibitemShut {NoStop}%
\bibitem [{\citenamefont {Rokhsari}\ \emph {et~al.}(2002)\citenamefont {Rokhsari}, \citenamefont {Kippenberg}, \citenamefont {Carmon}, \citenamefont {Vahala}, \citenamefont {Mancini}, \citenamefont {Giovanetti}, \citenamefont {Vitali}, \citenamefont {Tombesi}, \citenamefont {Armani}, \citenamefont {Spillane},\ and\ \citenamefont {J}}]{Rokhsari2002}%
  \BibitemOpen
  \bibfield  {author} {\bibinfo {author} {\bibfnamefont {H.}~\bibnamefont {Rokhsari}}, \bibinfo {author} {\bibfnamefont {T.~J.}\ \bibnamefont {Kippenberg}}, \bibinfo {author} {\bibfnamefont {T.}~\bibnamefont {Carmon}}, \bibinfo {author} {\bibfnamefont {K.~J.}\ \bibnamefont {Vahala}}, \bibinfo {author} {\bibfnamefont {S.}~\bibnamefont {Mancini}}, \bibinfo {author} {\bibfnamefont {V.}~\bibnamefont {Giovanetti}}, \bibinfo {author} {\bibfnamefont {D.}~\bibnamefont {Vitali}}, \bibinfo {author} {\bibfnamefont {P.}~\bibnamefont {Tombesi}}, \bibinfo {author} {\bibfnamefont {D.~K.}\ \bibnamefont {Armani}}, \bibinfo {author} {\bibfnamefont {S.~M.}\ \bibnamefont {Spillane}},\ and\ \bibinfo {author} {\bibfnamefont {K.~J. V.~K.}\ \bibnamefont {J}},\ }\href {https://doi.org/10.1364/OA_License_v1#VOR} {\bibfield  {journal} {\bibinfo  {journal} {Phys. Rev. Lett}\ }\textbf {\bibinfo {volume} {88}},\ \bibinfo {pages} {331} (\bibinfo {year} {2002})}\BibitemShut {NoStop}%
\bibitem [{\citenamefont {Schliesser}\ \emph {et~al.}(2008)\citenamefont {Schliesser}, \citenamefont {Rivière}, \citenamefont {Anetsberger}, \citenamefont {Arcizet},\ and\ \citenamefont {Kippenberg}}]{Schliesser2008}%
  \BibitemOpen
  \bibfield  {author} {\bibinfo {author} {\bibfnamefont {A.}~\bibnamefont {Schliesser}}, \bibinfo {author} {\bibfnamefont {R.}~\bibnamefont {Rivière}}, \bibinfo {author} {\bibfnamefont {G.}~\bibnamefont {Anetsberger}}, \bibinfo {author} {\bibfnamefont {O.}~\bibnamefont {Arcizet}},\ and\ \bibinfo {author} {\bibfnamefont {T.~J.}\ \bibnamefont {Kippenberg}},\ }\href {https://doi.org/10.1038/nphys939} {\bibfield  {journal} {\bibinfo  {journal} {Nature Physics}\ }\textbf {\bibinfo {volume} {4}},\ \bibinfo {pages} {415} (\bibinfo {year} {2008})}\BibitemShut {NoStop}%
\bibitem [{\citenamefont {Clarke}\ \emph {et~al.}(2023)\citenamefont {Clarke}, \citenamefont {Neveu}, \citenamefont {Khosla}, \citenamefont {Verhagen},\ and\ \citenamefont {Vanner}}]{Clarke2023}%
  \BibitemOpen
  \bibfield  {author} {\bibinfo {author} {\bibfnamefont {J.}~\bibnamefont {Clarke}}, \bibinfo {author} {\bibfnamefont {P.}~\bibnamefont {Neveu}}, \bibinfo {author} {\bibfnamefont {K.~E.}\ \bibnamefont {Khosla}}, \bibinfo {author} {\bibfnamefont {E.}~\bibnamefont {Verhagen}},\ and\ \bibinfo {author} {\bibfnamefont {M.~R.}\ \bibnamefont {Vanner}},\ }\bibfield  {journal} {\bibinfo  {journal} {Physical Review Letters}\ }\textbf {\bibinfo {volume} {131}},\ \href {https://doi.org/10.1103/PhysRevLett.131.053601} {10.1103/PhysRevLett.131.053601} (\bibinfo {year} {2023})\BibitemShut {NoStop}%
\bibitem [{\citenamefont {Weiss}\ \emph {et~al.}(2016)\citenamefont {Weiss}, \citenamefont {Kronwald},\ and\ \citenamefont {Marquardt}}]{weiss2016noise}%
  \BibitemOpen
  \bibfield  {author} {\bibinfo {author} {\bibfnamefont {T.}~\bibnamefont {Weiss}}, \bibinfo {author} {\bibfnamefont {A.}~\bibnamefont {Kronwald}},\ and\ \bibinfo {author} {\bibfnamefont {F.}~\bibnamefont {Marquardt}},\ }\href@noop {} {\bibfield  {journal} {\bibinfo  {journal} {New Journal of Physics}\ }\textbf {\bibinfo {volume} {18}},\ \bibinfo {pages} {013043} (\bibinfo {year} {2016})}\BibitemShut {NoStop}%
\bibitem [{\citenamefont {Catalini}\ \emph {et~al.}(2021)\citenamefont {Catalini}, \citenamefont {Rossi}, \citenamefont {Langman},\ and\ \citenamefont {Schliesser}}]{catalini2021modeling}%
  \BibitemOpen
  \bibfield  {author} {\bibinfo {author} {\bibfnamefont {L.}~\bibnamefont {Catalini}}, \bibinfo {author} {\bibfnamefont {M.}~\bibnamefont {Rossi}}, \bibinfo {author} {\bibfnamefont {E.~C.}\ \bibnamefont {Langman}},\ and\ \bibinfo {author} {\bibfnamefont {A.}~\bibnamefont {Schliesser}},\ }\href@noop {} {\bibfield  {journal} {\bibinfo  {journal} {Physical Review Letters}\ }\textbf {\bibinfo {volume} {126}},\ \bibinfo {pages} {174101} (\bibinfo {year} {2021})}\BibitemShut {NoStop}%
\bibitem [{\citenamefont {Parsa}\ \emph {et~al.}(2026)\citenamefont {Parsa}, \citenamefont {El-Sayed}, \citenamefont {Behjat}, \citenamefont {Barzanjeh},\ and\ \citenamefont {Barclay}}]{parsa2026large}%
  \BibitemOpen
  \bibfield  {author} {\bibinfo {author} {\bibfnamefont {P.}~\bibnamefont {Parsa}}, \bibinfo {author} {\bibfnamefont {W.}~\bibnamefont {El-Sayed}}, \bibinfo {author} {\bibfnamefont {P.}~\bibnamefont {Behjat}}, \bibinfo {author} {\bibfnamefont {S.}~\bibnamefont {Barzanjeh}},\ and\ \bibinfo {author} {\bibfnamefont {P.~E.}\ \bibnamefont {Barclay}},\ }\href@noop {} {\bibfield  {journal} {\bibinfo  {journal} {arXiv preprint arXiv:2603.19421}\ } (\bibinfo {year} {2026})}\BibitemShut {NoStop}%
\bibitem [{\citenamefont {Parsa}(2023)}]{Parsa2023}%
  \BibitemOpen
  \bibfield  {author} {\bibinfo {author} {\bibfnamefont {P.}~\bibnamefont {Parsa}},\ }\emph {\bibinfo {title} {Nonlinear Cavity Optomechanics in Diamond}},\ \href@noop {} {Master's thesis},\ \bibinfo  {school} {University of Calgary} (\bibinfo {year} {2023})\BibitemShut {NoStop}%
\bibitem [{\citenamefont {Meenehan}\ \emph {et~al.}(2015)\citenamefont {Meenehan}, \citenamefont {Cohen}, \citenamefont {MacCabe}, \citenamefont {Marsili}, \citenamefont {Shaw},\ and\ \citenamefont {Painter}}]{meenehan2015pulsed}%
  \BibitemOpen
  \bibfield  {author} {\bibinfo {author} {\bibfnamefont {S.~M.}\ \bibnamefont {Meenehan}}, \bibinfo {author} {\bibfnamefont {J.~D.}\ \bibnamefont {Cohen}}, \bibinfo {author} {\bibfnamefont {G.~S.}\ \bibnamefont {MacCabe}}, \bibinfo {author} {\bibfnamefont {F.}~\bibnamefont {Marsili}}, \bibinfo {author} {\bibfnamefont {M.~D.}\ \bibnamefont {Shaw}},\ and\ \bibinfo {author} {\bibfnamefont {O.}~\bibnamefont {Painter}},\ }\href@noop {} {\bibfield  {journal} {\bibinfo  {journal} {Physical Review X}\ }\textbf {\bibinfo {volume} {5}},\ \bibinfo {pages} {041002} (\bibinfo {year} {2015})}\BibitemShut {NoStop}%
\bibitem [{\citenamefont {Krause}\ \emph {et~al.}(2015)\citenamefont {Krause}, \citenamefont {Hill}, \citenamefont {Ludwig}, \citenamefont {Safavi-Naeini}, \citenamefont {Chan}, \citenamefont {Marquardt},\ and\ \citenamefont {Painter}}]{krause2015nonlinear}%
  \BibitemOpen
  \bibfield  {author} {\bibinfo {author} {\bibfnamefont {A.~G.}\ \bibnamefont {Krause}}, \bibinfo {author} {\bibfnamefont {J.~T.}\ \bibnamefont {Hill}}, \bibinfo {author} {\bibfnamefont {M.}~\bibnamefont {Ludwig}}, \bibinfo {author} {\bibfnamefont {A.~H.}\ \bibnamefont {Safavi-Naeini}}, \bibinfo {author} {\bibfnamefont {J.}~\bibnamefont {Chan}}, \bibinfo {author} {\bibfnamefont {F.}~\bibnamefont {Marquardt}},\ and\ \bibinfo {author} {\bibfnamefont {O.}~\bibnamefont {Painter}},\ }\href@noop {} {\bibfield  {journal} {\bibinfo  {journal} {Physical Review Letters}\ }\textbf {\bibinfo {volume} {115}},\ \bibinfo {pages} {233601} (\bibinfo {year} {2015})}\BibitemShut {NoStop}%
\bibitem [{\citenamefont {MacQuarrie}\ \emph {et~al.}(2015)\citenamefont {MacQuarrie}, \citenamefont {Gosavi}, \citenamefont {Moehle}, \citenamefont {Jungwirth}, \citenamefont {Bhave},\ and\ \citenamefont {Fuchs}}]{macquarrie2015coherent}%
  \BibitemOpen
  \bibfield  {author} {\bibinfo {author} {\bibfnamefont {E.}~\bibnamefont {MacQuarrie}}, \bibinfo {author} {\bibfnamefont {T.}~\bibnamefont {Gosavi}}, \bibinfo {author} {\bibfnamefont {A.}~\bibnamefont {Moehle}}, \bibinfo {author} {\bibfnamefont {N.}~\bibnamefont {Jungwirth}}, \bibinfo {author} {\bibfnamefont {S.}~\bibnamefont {Bhave}},\ and\ \bibinfo {author} {\bibfnamefont {G.}~\bibnamefont {Fuchs}},\ }\href@noop {} {\bibfield  {journal} {\bibinfo  {journal} {Optica}\ }\textbf {\bibinfo {volume} {2}},\ \bibinfo {pages} {233} (\bibinfo {year} {2015})}\BibitemShut {NoStop}%
\bibitem [{\citenamefont {Hu}\ \emph {et~al.}(2021)\citenamefont {Hu}, \citenamefont {Ding}, \citenamefont {Qin}, \citenamefont {Gu}, \citenamefont {Wan}, \citenamefont {Xiao},\ and\ \citenamefont {Jiang}}]{Hu2021}%
  \BibitemOpen
  \bibfield  {author} {\bibinfo {author} {\bibfnamefont {Y.}~\bibnamefont {Hu}}, \bibinfo {author} {\bibfnamefont {S.}~\bibnamefont {Ding}}, \bibinfo {author} {\bibfnamefont {Y.}~\bibnamefont {Qin}}, \bibinfo {author} {\bibfnamefont {J.}~\bibnamefont {Gu}}, \bibinfo {author} {\bibfnamefont {W.}~\bibnamefont {Wan}}, \bibinfo {author} {\bibfnamefont {M.}~\bibnamefont {Xiao}},\ and\ \bibinfo {author} {\bibfnamefont {X.}~\bibnamefont {Jiang}},\ }\bibfield  {journal} {\bibinfo  {journal} {Physical Review Letters}\ }\textbf {\bibinfo {volume} {127}},\ \href {https://doi.org/10.1103/PhysRevLett.127.134301} {10.1103/PhysRevLett.127.134301} (\bibinfo {year} {2021})\BibitemShut {NoStop}%
\bibitem [{\citenamefont {Gou}\ \emph {et~al.}(2025)\citenamefont {Gou}, \citenamefont {Privratsky}, \citenamefont {Sun}, \citenamefont {Liu}, \citenamefont {Abiri},\ and\ \citenamefont {Li}}]{gou2025chip}%
  \BibitemOpen
  \bibfield  {author} {\bibinfo {author} {\bibfnamefont {X.}~\bibnamefont {Gou}}, \bibinfo {author} {\bibfnamefont {W.}~\bibnamefont {Privratsky}}, \bibinfo {author} {\bibfnamefont {W.}~\bibnamefont {Sun}}, \bibinfo {author} {\bibfnamefont {Y.}~\bibnamefont {Liu}}, \bibinfo {author} {\bibfnamefont {H.}~\bibnamefont {Abiri}},\ and\ \bibinfo {author} {\bibfnamefont {Q.}~\bibnamefont {Li}},\ }\href@noop {} {\bibfield  {journal} {\bibinfo  {journal} {Nano Letters}\ }\textbf {\bibinfo {volume} {25}},\ \bibinfo {pages} {17644} (\bibinfo {year} {2025})}\BibitemShut {NoStop}%
\bibitem [{\citenamefont {Wang}\ \emph {et~al.}(2024{\natexlab{a}})\citenamefont {Wang}, \citenamefont {Hu}, \citenamefont {Lao}, \citenamefont {Wang}, \citenamefont {Jin}, \citenamefont {Zhou}, \citenamefont {Lei}, \citenamefont {Wang}, \citenamefont {Liu}, \citenamefont {Yang},\ and\ \citenamefont {Li}}]{Wang2024}%
  \BibitemOpen
  \bibfield  {author} {\bibinfo {author} {\bibfnamefont {M.}~\bibnamefont {Wang}}, \bibinfo {author} {\bibfnamefont {Z.~G.}\ \bibnamefont {Hu}}, \bibinfo {author} {\bibfnamefont {C.}~\bibnamefont {Lao}}, \bibinfo {author} {\bibfnamefont {Y.}~\bibnamefont {Wang}}, \bibinfo {author} {\bibfnamefont {X.}~\bibnamefont {Jin}}, \bibinfo {author} {\bibfnamefont {X.}~\bibnamefont {Zhou}}, \bibinfo {author} {\bibfnamefont {Y.}~\bibnamefont {Lei}}, \bibinfo {author} {\bibfnamefont {Z.}~\bibnamefont {Wang}}, \bibinfo {author} {\bibfnamefont {W.}~\bibnamefont {Liu}}, \bibinfo {author} {\bibfnamefont {Q.~F.}\ \bibnamefont {Yang}},\ and\ \bibinfo {author} {\bibfnamefont {B.~B.}\ \bibnamefont {Li}},\ }\bibfield  {journal} {\bibinfo  {journal} {Physical Review X}\ }\textbf {\bibinfo {volume} {14}},\ \href {https://doi.org/10.1103/PhysRevX.14.011056} {10.1103/PhysRevX.14.011056} (\bibinfo {year} {2024}{\natexlab{a}})\BibitemShut {NoStop}%
\bibitem [{\citenamefont {Ng}\ \emph {et~al.}(2023)\citenamefont {Ng}, \citenamefont {Nizet}, \citenamefont {Navarro-Urrios}, \citenamefont {Arregui}, \citenamefont {Albrechtsen}, \citenamefont {García}, \citenamefont {Stobbe}, \citenamefont {Sotomayor-Torres},\ and\ \citenamefont {Madiot}}]{Ng2023}%
  \BibitemOpen
  \bibfield  {author} {\bibinfo {author} {\bibfnamefont {R.~C.}\ \bibnamefont {Ng}}, \bibinfo {author} {\bibfnamefont {P.}~\bibnamefont {Nizet}}, \bibinfo {author} {\bibfnamefont {D.}~\bibnamefont {Navarro-Urrios}}, \bibinfo {author} {\bibfnamefont {G.}~\bibnamefont {Arregui}}, \bibinfo {author} {\bibfnamefont {M.}~\bibnamefont {Albrechtsen}}, \bibinfo {author} {\bibfnamefont {P.~D.}\ \bibnamefont {García}}, \bibinfo {author} {\bibfnamefont {S.}~\bibnamefont {Stobbe}}, \bibinfo {author} {\bibfnamefont {C.~M.}\ \bibnamefont {Sotomayor-Torres}},\ and\ \bibinfo {author} {\bibfnamefont {G.}~\bibnamefont {Madiot}},\ }\bibfield  {journal} {\bibinfo  {journal} {Physical Review Research}\ }\textbf {\bibinfo {volume} {5}},\ \href {https://doi.org/10.1103/PhysRevResearch.5.L032028} {10.1103/PhysRevResearch.5.L032028} (\bibinfo {year} {2023})\BibitemShut {NoStop}%
\bibitem [{\citenamefont {Wang}\ \emph {et~al.}(2024{\natexlab{b}})\citenamefont {Wang}, \citenamefont {Zhang}, \citenamefont {Shen}, \citenamefont {Xu}, \citenamefont {Niu}, \citenamefont {Sun}, \citenamefont {Guo},\ and\ \citenamefont {Dong}}]{wang2024optomechanical}%
  \BibitemOpen
  \bibfield  {author} {\bibinfo {author} {\bibfnamefont {Y.}~\bibnamefont {Wang}}, \bibinfo {author} {\bibfnamefont {M.}~\bibnamefont {Zhang}}, \bibinfo {author} {\bibfnamefont {Z.}~\bibnamefont {Shen}}, \bibinfo {author} {\bibfnamefont {G.-T.}\ \bibnamefont {Xu}}, \bibinfo {author} {\bibfnamefont {R.}~\bibnamefont {Niu}}, \bibinfo {author} {\bibfnamefont {F.-W.}\ \bibnamefont {Sun}}, \bibinfo {author} {\bibfnamefont {G.-C.}\ \bibnamefont {Guo}},\ and\ \bibinfo {author} {\bibfnamefont {C.-H.}\ \bibnamefont {Dong}},\ }\href@noop {} {\bibfield  {journal} {\bibinfo  {journal} {Physical Review Letters}\ }\textbf {\bibinfo {volume} {132}},\ \bibinfo {pages} {163603} (\bibinfo {year} {2024}{\natexlab{b}})}\BibitemShut {NoStop}%
\bibitem [{\citenamefont {Wan}\ \emph {et~al.}(2025)\citenamefont {Wan}, \citenamefont {Xia}, \citenamefont {Sun}, \citenamefont {Guan},\ and\ \citenamefont {Zhou}}]{wan2025optomechanical}%
  \BibitemOpen
  \bibfield  {author} {\bibinfo {author} {\bibfnamefont {X.}~\bibnamefont {Wan}}, \bibinfo {author} {\bibfnamefont {J.}~\bibnamefont {Xia}}, \bibinfo {author} {\bibfnamefont {H.}~\bibnamefont {Sun}}, \bibinfo {author} {\bibfnamefont {Q.}~\bibnamefont {Guan}},\ and\ \bibinfo {author} {\bibfnamefont {G.}~\bibnamefont {Zhou}},\ }\href@noop {} {\bibfield  {journal} {\bibinfo  {journal} {Communications Physics}\ }\textbf {\bibinfo {volume} {8}},\ \bibinfo {pages} {290} (\bibinfo {year} {2025})}\BibitemShut {NoStop}%
\bibitem [{\citenamefont {MacCabe}\ \emph {et~al.}(2020)\citenamefont {MacCabe}, \citenamefont {Ren}, \citenamefont {Luo}, \citenamefont {Cohen}, \citenamefont {Zhou}, \citenamefont {Sipahigil}, \citenamefont {Mirhosseini},\ and\ \citenamefont {Painter}}]{maccabe2020nano}%
  \BibitemOpen
  \bibfield  {author} {\bibinfo {author} {\bibfnamefont {G.~S.}\ \bibnamefont {MacCabe}}, \bibinfo {author} {\bibfnamefont {H.}~\bibnamefont {Ren}}, \bibinfo {author} {\bibfnamefont {J.}~\bibnamefont {Luo}}, \bibinfo {author} {\bibfnamefont {J.~D.}\ \bibnamefont {Cohen}}, \bibinfo {author} {\bibfnamefont {H.}~\bibnamefont {Zhou}}, \bibinfo {author} {\bibfnamefont {A.}~\bibnamefont {Sipahigil}}, \bibinfo {author} {\bibfnamefont {M.}~\bibnamefont {Mirhosseini}},\ and\ \bibinfo {author} {\bibfnamefont {O.}~\bibnamefont {Painter}},\ }\href@noop {} {\bibfield  {journal} {\bibinfo  {journal} {Science}\ }\textbf {\bibinfo {volume} {370}},\ \bibinfo {pages} {840} (\bibinfo {year} {2020})}\BibitemShut {NoStop}%
\bibitem [{\citenamefont {Li}\ \emph {et~al.}(2024)\citenamefont {Li}, \citenamefont {Lekavicius}, \citenamefont {Noeckel},\ and\ \citenamefont {Wang}}]{li2024ultracoherent}%
  \BibitemOpen
  \bibfield  {author} {\bibinfo {author} {\bibfnamefont {X.}~\bibnamefont {Li}}, \bibinfo {author} {\bibfnamefont {I.}~\bibnamefont {Lekavicius}}, \bibinfo {author} {\bibfnamefont {J.}~\bibnamefont {Noeckel}},\ and\ \bibinfo {author} {\bibfnamefont {H.}~\bibnamefont {Wang}},\ }\href@noop {} {\bibfield  {journal} {\bibinfo  {journal} {Nano Letters}\ }\textbf {\bibinfo {volume} {24}},\ \bibinfo {pages} {10995} (\bibinfo {year} {2024})}\BibitemShut {NoStop}%
\bibitem [{\citenamefont {Wallucks}\ \emph {et~al.}(2020)\citenamefont {Wallucks}, \citenamefont {Marinkovi{\'c}}, \citenamefont {Hensen}, \citenamefont {Stockill},\ and\ \citenamefont {Gr{\"o}blacher}}]{wallucks2020quantum}%
  \BibitemOpen
  \bibfield  {author} {\bibinfo {author} {\bibfnamefont {A.}~\bibnamefont {Wallucks}}, \bibinfo {author} {\bibfnamefont {I.}~\bibnamefont {Marinkovi{\'c}}}, \bibinfo {author} {\bibfnamefont {B.}~\bibnamefont {Hensen}}, \bibinfo {author} {\bibfnamefont {R.}~\bibnamefont {Stockill}},\ and\ \bibinfo {author} {\bibfnamefont {S.}~\bibnamefont {Gr{\"o}blacher}},\ }\href@noop {} {\bibfield  {journal} {\bibinfo  {journal} {Nature Physics}\ }\textbf {\bibinfo {volume} {16}},\ \bibinfo {pages} {772} (\bibinfo {year} {2020})}\BibitemShut {NoStop}%
\bibitem [{\citenamefont {Kristensen}\ \emph {et~al.}(2024)\citenamefont {Kristensen}, \citenamefont {Kralj}, \citenamefont {Langman},\ and\ \citenamefont {Schliesser}}]{kristensen2024long}%
  \BibitemOpen
  \bibfield  {author} {\bibinfo {author} {\bibfnamefont {M.~B.}\ \bibnamefont {Kristensen}}, \bibinfo {author} {\bibfnamefont {N.}~\bibnamefont {Kralj}}, \bibinfo {author} {\bibfnamefont {E.~C.}\ \bibnamefont {Langman}},\ and\ \bibinfo {author} {\bibfnamefont {A.}~\bibnamefont {Schliesser}},\ }\href@noop {} {\bibfield  {journal} {\bibinfo  {journal} {Physical review letters}\ }\textbf {\bibinfo {volume} {132}},\ \bibinfo {pages} {100802} (\bibinfo {year} {2024})}\BibitemShut {NoStop}%
\bibitem [{\citenamefont {Stannigel}\ \emph {et~al.}(2010)\citenamefont {Stannigel}, \citenamefont {Rabl}, \citenamefont {S\o{}rensen}, \citenamefont {Zoller},\ and\ \citenamefont {Lukin}}]{Stannigel2010transduction}%
  \BibitemOpen
  \bibfield  {author} {\bibinfo {author} {\bibfnamefont {K.}~\bibnamefont {Stannigel}}, \bibinfo {author} {\bibfnamefont {P.}~\bibnamefont {Rabl}}, \bibinfo {author} {\bibfnamefont {A.~S.}\ \bibnamefont {S\o{}rensen}}, \bibinfo {author} {\bibfnamefont {P.}~\bibnamefont {Zoller}},\ and\ \bibinfo {author} {\bibfnamefont {M.~D.}\ \bibnamefont {Lukin}},\ }\href {https://doi.org/10.1103/PhysRevLett.105.220501} {\bibfield  {journal} {\bibinfo  {journal} {Phys. Rev. Lett.}\ }\textbf {\bibinfo {volume} {105}},\ \bibinfo {pages} {220501} (\bibinfo {year} {2010})}\BibitemShut {NoStop}%
\bibitem [{\citenamefont {Sonar}\ \emph {et~al.}(2025)\citenamefont {Sonar}, \citenamefont {Hatipoglu}, \citenamefont {Meesala}, \citenamefont {Lake}, \citenamefont {Ren},\ and\ \citenamefont {Painter}}]{sonar2025high}%
  \BibitemOpen
  \bibfield  {author} {\bibinfo {author} {\bibfnamefont {S.}~\bibnamefont {Sonar}}, \bibinfo {author} {\bibfnamefont {U.}~\bibnamefont {Hatipoglu}}, \bibinfo {author} {\bibfnamefont {S.}~\bibnamefont {Meesala}}, \bibinfo {author} {\bibfnamefont {D.~P.}\ \bibnamefont {Lake}}, \bibinfo {author} {\bibfnamefont {H.}~\bibnamefont {Ren}},\ and\ \bibinfo {author} {\bibfnamefont {O.}~\bibnamefont {Painter}},\ }\href@noop {} {\bibfield  {journal} {\bibinfo  {journal} {Optica}\ }\textbf {\bibinfo {volume} {12}},\ \bibinfo {pages} {99} (\bibinfo {year} {2025})}\BibitemShut {NoStop}%
\end{thebibliography}%


\begin{thebibliography}{4}%
\makeatletter
\providecommand \@ifxundefined [1]{%
 \@ifx{#1\undefined}
}%
\providecommand \@ifnum [1]{%
 \ifnum #1\expandafter \@firstoftwo
 \else \expandafter \@secondoftwo
 \fi
}%
\providecommand \@ifx [1]{%
 \ifx #1\expandafter \@firstoftwo
 \else \expandafter \@secondoftwo
 \fi
}%
\providecommand \natexlab [1]{#1}%
\providecommand \enquote  [1]{``#1''}%
\providecommand \bibnamefont  [1]{#1}%
\providecommand \bibfnamefont [1]{#1}%
\providecommand \citenamefont [1]{#1}%
\providecommand \href@noop [0]{\@secondoftwo}%
\providecommand \href [0]{\begingroup \@sanitize@url \@href}%
\providecommand \@href[1]{\@@startlink{#1}\@@href}%
\providecommand \@@href[1]{\endgroup#1\@@endlink}%
\providecommand \@sanitize@url [0]{\catcode `\\12\catcode `\$12\catcode `\&12\catcode `\#12\catcode `\^12\catcode `\_12\catcode `\%12\relax}%
\providecommand \@@startlink[1]{}%
\providecommand \@@endlink[0]{}%
\providecommand \url  [0]{\begingroup\@sanitize@url \@url }%
\providecommand \@url [1]{\endgroup\@href {#1}{\urlprefix }}%
\providecommand \urlprefix  [0]{URL }%
\providecommand \Eprint [0]{\href }%
\providecommand \doibase [0]{https://doi.org/}%
\providecommand \selectlanguage [0]{\@gobble}%
\providecommand \bibinfo  [0]{\@secondoftwo}%
\providecommand \bibfield  [0]{\@secondoftwo}%
\providecommand \translation [1]{[#1]}%
\providecommand \BibitemOpen [0]{}%
\providecommand \bibitemStop [0]{}%
\providecommand \bibitemNoStop [0]{.\EOS\space}%
\providecommand \EOS [0]{\spacefactor3000\relax}%
\providecommand \BibitemShut  [1]{\csname bibitem#1\endcsname}%
\let\auto@bib@innerbib\@empty
\bibitem [{\citenamefont {Safavi-Naeini}\ and\ \citenamefont {Painter}(2010)}]{Safavi2010}%
  \BibitemOpen
  \bibfield  {author} {\bibinfo {author} {\bibfnamefont {A.~H.}\ \bibnamefont {Safavi-Naeini}}\ and\ \bibinfo {author} {\bibfnamefont {O.}~\bibnamefont {Painter}},\ }\href@noop {} {\bibfield  {journal} {\bibinfo  {journal} {Optics express}\ }\textbf {\bibinfo {volume} {18}},\ \bibinfo {pages} {14926} (\bibinfo {year} {2010})}\BibitemShut {NoStop}%
\bibitem [{\citenamefont {Chan}\ \emph {et~al.}(2011)\citenamefont {Chan}, \citenamefont {Alegre}, \citenamefont {Safavi-Naeini}, \citenamefont {Hill}, \citenamefont {Krause}, \citenamefont {Gröblacher}, \citenamefont {Aspelmeyer},\ and\ \citenamefont {Painter}}]{Chan2011}%
  \BibitemOpen
  \bibfield  {author} {\bibinfo {author} {\bibfnamefont {J.}~\bibnamefont {Chan}}, \bibinfo {author} {\bibfnamefont {T.~P.}\ \bibnamefont {Alegre}}, \bibinfo {author} {\bibfnamefont {A.~H.}\ \bibnamefont {Safavi-Naeini}}, \bibinfo {author} {\bibfnamefont {J.~T.}\ \bibnamefont {Hill}}, \bibinfo {author} {\bibfnamefont {A.}~\bibnamefont {Krause}}, \bibinfo {author} {\bibfnamefont {S.}~\bibnamefont {Gröblacher}}, \bibinfo {author} {\bibfnamefont {M.}~\bibnamefont {Aspelmeyer}},\ and\ \bibinfo {author} {\bibfnamefont {O.}~\bibnamefont {Painter}},\ }\href {https://doi.org/10.1038/nature10461} {\bibfield  {journal} {\bibinfo  {journal} {Nature}\ }\textbf {\bibinfo {volume} {478}},\ \bibinfo {pages} {89} (\bibinfo {year} {2011})}\BibitemShut {NoStop}%
\bibitem [{\citenamefont {Zohari}\ \emph {et~al.}(2022)\citenamefont {Zohari}, \citenamefont {Losby}, \citenamefont {El-Sayed}, \citenamefont {Behjat}, \citenamefont {Luiz}, \citenamefont {Davis},\ and\ \citenamefont {Barclay}}]{Zohari2022}%
  \BibitemOpen
  \bibfield  {author} {\bibinfo {author} {\bibfnamefont {E.}~\bibnamefont {Zohari}}, \bibinfo {author} {\bibfnamefont {J.~E.}\ \bibnamefont {Losby}}, \bibinfo {author} {\bibfnamefont {W.}~\bibnamefont {El-Sayed}}, \bibinfo {author} {\bibfnamefont {P.}~\bibnamefont {Behjat}}, \bibinfo {author} {\bibfnamefont {G.~D.~O.}\ \bibnamefont {Luiz}}, \bibinfo {author} {\bibfnamefont {J.~P.}\ \bibnamefont {Davis}},\ and\ \bibinfo {author} {\bibfnamefont {P.~E.}\ \bibnamefont {Barclay}},\ }in\ \href {https://doi.org/10.1109/IPC53466.2022.9975524} {\emph {\bibinfo {booktitle} {2022 IEEE Photonics Conference, IPC 2022 - Proceedings}}}\ (\bibinfo  {publisher} {Institute of Electrical and Electronics Engineers Inc.},\ \bibinfo {year} {2022})\BibitemShut {NoStop}%
\bibitem [{\citenamefont {El-Sayed}\ \emph {et~al.}(2026)\citenamefont {El-Sayed}, \citenamefont {Zohari}, \citenamefont {Itoi}, \citenamefont {Parsa}, \citenamefont {Luiz}, \citenamefont {Losby}, \citenamefont {Hayashida}, \citenamefont {Malac},\ and\ \citenamefont {Barclay}}]{elsayed2026exceptional}%
  \BibitemOpen
  \bibfield  {author} {\bibinfo {author} {\bibfnamefont {W.}~\bibnamefont {El-Sayed}}, \bibinfo {author} {\bibfnamefont {E.}~\bibnamefont {Zohari}}, \bibinfo {author} {\bibfnamefont {J.}~\bibnamefont {Itoi}}, \bibinfo {author} {\bibfnamefont {P.}~\bibnamefont {Parsa}}, \bibinfo {author} {\bibfnamefont {G.~d.~O.}\ \bibnamefont {Luiz}}, \bibinfo {author} {\bibfnamefont {J.~E.}\ \bibnamefont {Losby}}, \bibinfo {author} {\bibfnamefont {M.}~\bibnamefont {Hayashida}}, \bibinfo {author} {\bibfnamefont {M.}~\bibnamefont {Malac}},\ and\ \bibinfo {author} {\bibfnamefont {P.~E.}\ \bibnamefont {Barclay}},\ }\href@noop {} {\bibfield  {journal} {\bibinfo  {journal} {arXiv preprint arXiv:2605.27536}\ } (\bibinfo {year} {2026})}\BibitemShut {NoStop}%
\end{thebibliography}%


\begin{thebibliography}{0}%
\makeatletter
\providecommand \@ifxundefined [1]{%
 \@ifx{#1\undefined}
}%
\providecommand \@ifnum [1]{%
 \ifnum #1\expandafter \@firstoftwo
 \else \expandafter \@secondoftwo
 \fi
}%
\providecommand \@ifx [1]{%
 \ifx #1\expandafter \@firstoftwo
 \else \expandafter \@secondoftwo
 \fi
}%
\providecommand \natexlab [1]{#1}%
\providecommand \enquote  [1]{``#1''}%
\providecommand \bibnamefont  [1]{#1}%
\providecommand \bibfnamefont [1]{#1}%
\providecommand \citenamefont [1]{#1}%
\providecommand \href@noop [0]{\@secondoftwo}%
\providecommand \href [0]{\begingroup \@sanitize@url \@href}%
\providecommand \@href[1]{\@@startlink{#1}\@@href}%
\providecommand \@@href[1]{\endgroup#1\@@endlink}%
\providecommand \@sanitize@url [0]{\catcode `\\12\catcode `\$12\catcode `\&12\catcode `\#12\catcode `\^12\catcode `\_12\catcode `\%12\relax}%
\providecommand \@@startlink[1]{}%
\providecommand \@@endlink[0]{}%
\providecommand \url  [0]{\begingroup\@sanitize@url \@url }%
\providecommand \@url [1]{\endgroup\@href {#1}{\urlprefix }}%
\providecommand \urlprefix  [0]{URL }%
\providecommand \Eprint [0]{\href }%
\providecommand \doibase [0]{https://doi.org/}%
\providecommand \selectlanguage [0]{\@gobble}%
\providecommand \bibinfo  [0]{\@secondoftwo}%
\providecommand \bibfield  [0]{\@secondoftwo}%
\providecommand \translation [1]{[#1]}%
\providecommand \BibitemOpen [0]{}%
\providecommand \bibitemStop [0]{}%
\providecommand \bibitemNoStop [0]{.\EOS\space}%
\providecommand \EOS [0]{\spacefactor3000\relax}%
\providecommand \BibitemShut  [1]{\csname bibitem#1\endcsname}%
\let\auto@bib@innerbib\@empty
\end{thebibliography}%
\end{document}